\documentclass[structabstract]{aa}  
\usepackage{graphicx,lscape,longtable}
\usepackage{txfonts,natbib}

\bibpunct{(}{)}{;}{a}{}{,}
\begin{document}
\title{Present-day cosmic abundances}   
\subtitle{A comprehensive study of nearby early B-type stars and
implications for\\ stellar and Galactic evolution, and interstellar dust models\thanks{Based on observations collected at the
   Centro Astron\'omico His\-pano Alem\'an (CAHA) at Calar Alto, operated 
   jointly by the Max-
   Planck Institut f\"ur Astronomie and the Instituto de Astrof\'isica
   de Andaluc\'ia (CSIC), proposals H2001-2.2-011
   and H2005-2.2-016.}\fnmsep
   \thanks{Based on observations collected at the European Southern
   Obser\-vatory, Chile, ESO 074.B-0455(A).}\fnmsep
   \thanks{Based on spectral data retrieved from the ELODIE archive at
   Ob\-servatoire de Haute-Provence (OHP).}}
   
\author{Mar\'ia-Fernanda Nieva\inst{1,2}
          \and
          Norbert Przybilla\inst{2}
          }

   \institute{Max-Planck-Institut f\"ur Astrophysik,
   Karl-Schwarzschild-Str.~1, D-85741 Garching, Germany
   \and
   Dr. Karl Remeis-Sternwarte \& ECAP, Universit\"at Erlangen-N\"urnberg, 
   Sternwartstr. 7, D-96049 Bamberg, Germany\\
   \email{Maria-Fernanda.Nieva@sternwarte.uni-erlangen.de;Norbert.Przybilla@sternwarte.uni-erlangen.de}
             }

   \date{}

  \abstract
  {}
   {A carefully selected sample of early B-type stars in OB
   associations and the field within the solar neighbourhood 
   is studied comprehensively. 
    Present-day abundances for the astrophysically most 
    interesting chemical elements are derived in order to investigate
    whether a present-day cosmic abundance standard can be established.}
   {High-resolution and high-S/N spectra of 
   well-studied sharp-lined early B-type stars 
   are analysed in non-LTE. 
   Atmospheric parameters are derived from the
   simultaneous establishment of independent indicators, from
   multiple ionization equilibria and the hydrogen Balmer lines, and they are confirmed by
   reproduction of the stars' global spectral energy~distributions.}
   {Effective temperatures are constrained to 1-2\% and
   surface gravities to less than 15\% uncertainty, along with
   accurate rotational, micro- and macroturbulence velocities.
   Good agreement of the resulting spectroscopic parallaxes with
   those from the new reduction of the Hipparcos catalogue is obtained.  
   Absolute values for abundances of He, C, N, O, Ne, Mg, Si 
   and Fe are determined to better than 25\% uncertainty. 
   The synthetic spectra match the observations reliably over almost the
   entire visual spectral~range.}
   {A present-day cosmic abundance standard is established from 
   a sample of 29 early B-type stars, indicating abundance fluctuations
   of less than 10\% around the mean. Our results
   ({\sc i}) resolve the long-standing discrepancy 
   between a chemical homogeneous gas-phase ISM and a chemically
   inhomogeneous young stellar component out to several hundred parsec 
   from the Sun,
   ({\sc ii}) facilitate the amount of heavy elements locked up in the 
   interstellar dust to be constrained precisely -- the results
   imply that carbonaceous dust is largely destroyed inside the 
   Orion \ion{H}{ii} region, unlike the silicates, and that graphite
   is only a minority species in interstellar dust --,
   ({\sc iii}) show that the mixing of CNO-burning products in the
   course of massive star evolution follows tightly the predicted nuclear path,
   ({\sc iv}) provide reliable present-day reference points for
   anchoring Galactic chemical evolution models to observation, and
   ({\sc v}) imply that the Sun has migrated outwards
   from the inner Galactic disk over its lifetime from a
   birthplace at a distance around 5-6\,kpc from the Galactic Centre; 
   a cancellation of the effects of Galactic chemical evolution and
   abundance gradients leads to the similarity of solar and
   present-day cosmic abundances in the solar neighbourhood, with a
   telltaling signature of the Sun's origin left in the C/O ratio. (ABRIDGED) 
}
   

   \authorrunning{M.F. Nieva \& N. Przybilla}
   \titlerunning{Present-day cosmic abundances}
   
   \maketitle
%

\section{Introduction}

The formation and evolution of galaxies, stars, interstellar gas and dust, 
planetary systems, and even life are tightly related to the origin and 
evolution of the chemical elements, and therefore to the cosmic cycle 
of matter. Theories that aim at explaining these phenomena need to be
anchored to reference values for chemical abundances. 
The most appropriate source of reference abundances for the 
chemical composition of cosmic matter is a topic of vivid discussion.

The chemical composition of most astronomical objects is
traditionally compared to that of the Sun because it is the closest star and 
therefore its chemical abundances can be determined with high accuracy and precision
from application of various techniques to observations of very high quality.
The solar photosphere, its chromosphere and corona can be studied remotely
via spectroscopy. Particle collection techniques from space also
allow {\em in-situ} measurements of the chemical composition of the solar 
wind and of solar energetic particles. Moreover, mass spectroscopy of 
CI-chondrites facilitates the pristine composition of
the solar nebula to be accessed (except for a few volatile elements). 
The wealth of data gave rise to the solar abundance standard, which 
has been subject of active research in the past decades and 
it is still subject of continuous revision and improvement 
based on the modeling of its convective outer envelope and the radiative
transfer in 3D models \citep[e.g.][and references 
therein]{AGSS09,LPG09,Caffau10}.

Conceptually, the chemical composition of the Sun constitutes an excellent 
reference for numerous astrophysical studies, of e.g. low-mass stars' interiors and 
atmospheres, the Galactic chemical evolution of `older' star 
populations, or of solar twins (with and without planets), to mention only 
few among many other applications. In doing this, one supposes that
the Sun is a typical, middle-aged low-mass star. However, the concept faces
complications as different sets of 
reference values for the Sun are discussed in the literature. Objective 
criteria that allow one to decide which is the most appropriate on a 
level where the details matter do not exist, as each of them may 
reproduce one or another aspect of observations better\footnote{The
contemporarily most prominent example may be the so-called solar composition 
problem \citep{BaAn08}, the mismatch of solar interior models based on
modern (lower) solar abundances with helioseismic observations, whereas
consensus was established with previous higher metallicity values of
\citet{gs98}.}. Because of this we will refer to the generic solar values 
henceforth as the `solar standard' and specify individual sources
where relevant.

There are also examples where the Sun is considered as 
a reference because of the lack of another reliable standard, despite
its chemical composition may not be representative for the objects
of study, e.g. whenever a local and present-day reference is required.
This is because the Sun can {\em a priori} not be assumed to provide a
standard for the chemical composition at present-day. 
The reason for this is that Galactic chemical evolution has 
proceeded for the 4.56\,Gyrs since the formation of the Sun,
enriching the interstellar medium (ISM) in heavy elements. 
For instance, structure and evolution models of massive stars
anchor their initial composition to the solar standard despite these
young stars are formed from the present-day ISM. Similarly,
the solar standard is used to constrain the amount of heavy elements 
locked up in the dust-phase of the present-day ISM. Galactic chemical evolution
models also consider the solar chemical abundances as representative for 
the `local' Galactic neighbourhood of the Sun\footnote{The extent of the solar 
neighbourhood has different meanings depending on the context. For 
instance, in studies of F and G stars it is typically stretched out to 
several tens of pc, while here we consider it to extend out to 
$\sim$ 400\,pc, similar to studies of the ISM.}
while it was possibly formed nowhere near its current
Galactocentric position. Moreover, Galactic chemical evolution models anchor the 
present-day composition of the solar neighbourhood to the solar 
standard in aiming to reproduce present-day Galactic abundance gradients.
These are only a few among many other applications. 
Establishing an appropriate abundance standard for the present-day in
our local Galactic environment would be highly valuable either to verify
whether the assumption that the solar standard is indeed
representative, or to constrain various theories in the 
astrophysical context more tightly.

Present-day chemical abundances in the solar neighbourhood 
can be accessed relatively easily from absorption line studies of 
the cold/warm ISM or from emission line spectroscopy of the 
Orion nebula. In particular the \ion{H}{ii} regions are regarded as 
privileged sites for the determination of chemical abundances over large 
distances, even in other galaxies. 
Unfortunately, the composition of the neutral and the ionized 
ISM gas is altered by depletion onto dust grains. In order to access 
the actual chemical composition accurately, an a-priori knowledge
of the composition of the dust-phase is required, which, however, is lacking.
Further complications for \ion{H}{ii} region studies are the dependence
of the derived abundances on the indicators employed in the
analysis\footnote{Recombination or collisionally excited lines
can indicate rather discrepant results \citep[see e.g.][]{sergio11}.},
fluctuations of the electron temperature throughout the nebula, and 
ionization correction factors, which can be substantial for some 
chemical species.

An ideal alternative to find a reference for the 
chemical composition of cosmic matter 
are normal unevolved early B-type stars of $\sim$8-18\,$M_{\sun}$,
which can provide simultaneously temporal (present-day) and local (birth place) 
information on chemical abundances. They can be observed not only in the solar 
neighbourhood, but also at larger distances in the Milky Way
and in other galaxies.
Their composition is unaffected by depletion onto dust grains, 
unlike the cool/warm ISM and the \ion{H}{ii} regions.
In contrast to cooler and lower mass stars they are so short lived 
(typically $\sim$10$^7$\,yrs) that they have no time to travel far from 
their birth place, except for the occasional~runaway~star.
From an analysis perspective, spectroscopic studies of early B-type stars 
are relatively simple because their photospheres are not affected 
by strong stellar winds, unlike their hotter and more luminous siblings,
or by convection and chromospheres, unlike the cool stars, which
furthermore pose more challenges because of severe line blending in
their crowded optical spectra.
The atmospheres of unevolved early B-type stars are well represented by 
classical hydrostatic, plane-parallel
1D-models in radiative and local thermodynamic equilibrium
\citep[LTE;][henceforth abbreviated NP07]{np07} -- 
though deviations from LTE (non-LTE effects) need to be accounted for
{\em properly} in line-formation calculations \citep{pnb11}. Moreover, they
are typically unaffected by atomic diffusion that gives rise to peculiarities of
elemental abundances in many mid B- to A-type stars \citep[e.g.][]{Smith96}.
In slowly-rotating stars the photosphere should also be
essentially unaffected by rotational mixing with CNO-cycled matter
from the stellar core
(except for some fast rotators seen pole-on), i.e. they should retain
their pristine chemical surface composition throughout their
main-sequence phase.

\begin{figure*}[ht!]
\sidecaption
\includegraphics[width=12cm]{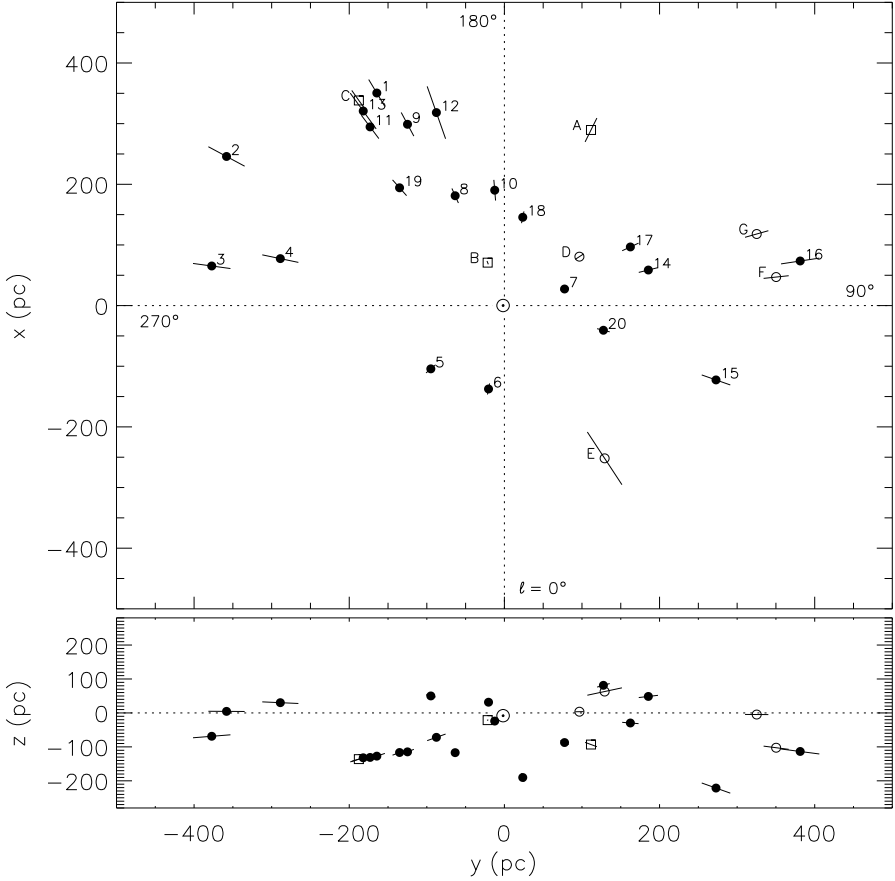}
\caption{Distribution of the sample stars in the solar neighbourhood. 
Upper panel: Galactic plane projection, the Galactic centre lies in 
direction of the bottom of the panel. Lower panel: rotational 
plane projection, the Northern Galactic Pole lies in direction of the 
top of the panel. The sample stars that are analysed quantitatively 
here are marked by dots, open boxes mark candidate SB2 
systems and open circles stars that were excluded from the analysis 
because of other peculiarities. Numbers/letters correspond to those of 
our internal numbering in the tables. 
Error bars are of spectroscopic distances (stars 1--20, see
Table~\ref{bigtable}) and of measured parallaxes (objects A--G). 
The position of the Sun is marked by $\sun$.}
\label{distribution}
\end{figure*}

Despite the relatively simple physics, spectral analyses of main 
sequence B-stars have turned out to provide inconclusive results over the past 
decades, i.e. large uncertainties in basic stellar parameters, 
a tendency towards a metal-poor composition with respect to older 
stars like the Sun 
and an overall enormous range in derived elemental abundances, challenging 
predictions of stellar and Galactic evolution models. For a 
discussion on these topics see e.g. the reviews of
\citet{Herrero03}, \citet{HeLe04}, \citet{p08r} and \citet{Morel09}.

In order to improve the quantitative analysis of this type of
stars we have 
extensively updated the spectral modelling by constructing robust 
model atoms for non-LTE line-formation calculations and implemented a
powerful self-consistent analysis technique, which brings numerous spectroscopic
parameter and abundance indicators into agreement simultaneously, 
starting with H, He and C \citep{n07}.
First applications of our method to carbon abundances in a small sample
of stars in the solar neighbourhood \citep[NP08]{np08} provided an unprecedented 
small scatter and an absolute average abundance similar to the solar
value by \citet{AGS05}, 
solving for the first time the above mentioned problems. 
Our efforts, also on other metals \citep[PNB08]{pnb08}, 
have provided highly-promising results so far, i.e. a drastic
reduction of statistical and systematic errors in stellar parameters and
chemical abundances: uncertainties as low as $\sim$1\% in effective 
temperature, $\sim$10\% in surface gravity and 
$\sim$10-20\% in 
elemental abundances have been achieved --- compared to 
$\sim$5-10\%, 
$\sim$25\% and a factor $\sim$2, respectively, using standard 
methods\footnote{Note that typical uncertainties for effective temperature 
given in the literature
are $\le$3-5\%, however these do not take into account the large
discrepancies -- up to $\sim$4000\,K -- that exist between 
different temperature determination methods, see Sect.~5.2 of NP08 
for a discussion. In the extremes, error estimates of up to 40-60\% in 
surface gravity and a factor 3-5 in elemental abundances are 
found in the literature \citep[e.g.][]{hunter07,trundle07}.}. 
The initial sample of 
6 stars from OB associations and the field in the solar neighbourhood
covering a broad parameter range turned out to be
chemically homogeneous on the 10\% level, 
corroborating earlier findings from analyses of the ISM gas-phase (PNB08). 
Yet, the sample size was arguably~small.

In this work we therefore present a comprehensive study of a sample of
initially 27 early B-type stars 
from the solar neighborhood, aiming to re-investigate our previous 
findings (from a sub-sample of only 6 stars) 
on a solid statistical basis.
The stars span about 20\,000\,K in effective temperature and 
$\sim$0.7\,dex in surface gravities, from the zero-age 
main sequence (ZAMS) to close to the end of the core hydrogen
burning phase. Seven stars from the original sample had to be excluded 
from the analysis for various reasons. In order to compensate for this
loss, nine additional stars from the Ori OB1 association analysed in an 
identical way \citep[NS11]{ns11} were included in the derivation of the proposed 
{\em cosmic abundance standard}.

The paper is organised as follows. Particular attention is paid 
to the selection criteria for the star sample and of 
the special care taken to obtain observations of high quality in
Sect.~\ref{sect_obs}. The computations of extensive grids of 
synthetic spectra and the implementation of a semi-automatic 
analysis technique are discussed in Sect.~\ref{sect_the}, which 
facilitates the same quality of the analyses to be achieved
as in our previous work (Sect.~\ref{sect_an}).
A comprehensive characterisation of the atmospheric properties of the
sample stars and related quantities is provided in Sect.~\ref{params}.
Then, the resulting present-day cosmic abundance standard for 
the solar neighbourhood is discussed in Sect.~\ref{sect_abund}. 
Our results are put in the broader astrophysical context in
Sect.~\ref{sect_impl}, concentrating on ISM science, and stellar 
and Galactic chemical evolution. Finally, a summary is given 
in~Sect.~\ref{summary}.


\begin{table*}
\centering
\caption[]{
The complete star sample: id, spectral type, variability, 
OB association membership, photometry\tablefootmark{a}, and observational details.\\[-6mm] \label{observations}}
\footnotesize
 \setlength{\tabcolsep}{.1cm}
 \begin{tabular}{r@{\hspace{1mm}}rrclcc@{\hspace{1mm}}rrrrrrrr}
 \noalign{}
\hline
\hline
\#  & HD  & HR  &Name & Sp.\,T\tablefootmark{b} & Variab. & OB\,Assoc.&& $V$ & $B-V$ & Date       & $N$ & $T_\mathrm{exp}$ & S/N$_B$ \\
  &     &     &     &              &         &           && mag &  mag  & DD\,MM\,YYYY &     & s                & \\
\hline\\[-3mm]
{\sc Feros}~&$R$\,=\,48\,000\\
\hline\\[-2mm]
            1 & 36591 & 1861&               &B1\,V   &             &Ori\,OB1\,Ib&    &  5.339 & $-$0.194  & 25-02-2005 & 1 &  120 & 530\\[-.2mm]
              &       &     &               &        &             &        &  $\pm$ &  ...   & $ $...    &\\[-.2mm]
            2 & 61068 & 2928&PT\,Pup        &B2\,III & $\beta$\,Cep& Field  &        &  5.711 & $-$0.176  & 25-02-2005 & 1 &  180 & 420\\[-.2mm]
              &       &     &               &        &             &        &  $\pm$ &  0.015 & $ $0.010  &\\[-.2mm]
            3 & 63922 & 3055&               &B0\,III &             & Field  &        &  4.106 & $-$0.185  & 25-02-2005 & 1 &   60 & 470\\[-.2mm]
              &       &     &               &        &             &        &  $\pm$ &  0.005 & $ $0.007  &\\[-.2mm]
            4 & 74575 & 3468&$\alpha$\,Pyx  &B1.5\,III&$\beta$\,Cep& Field  &        &  3.679 & $-$0.183  & 27-02-2005 & 2 &   60 & 300\\[-.2mm]
              &       &     &               &        &             &        &  $\pm$ &  0.006 & $ $0.003  &\\[-.2mm]
            5 &122980 & 5285&$\chi$\,Cen    &B2\,V   &$\beta$\,Cep &Up.\,Cen\,Lup&   &  4.353 & $-$0.195  & 26-02-2005 & 1 &   60 & 480\\[-.2mm]
              &       &     &               &        &             &        &  $\pm$ &  0.007 & $ $0.006  &\\[-.2mm]
            6 &149438 & 6165&$\tau$\,Sco    &B0.2\,V &             &Up.\,Sco&        &  2.825 & $-$0.252  & 24-02-2005 & 1 &   80 & 810\\[-.2mm]
              &       &     &               &        &             &        &  $\pm$ &  0.009 & $ $0.007  & 25-02-2005 & 3\\
\hline\\[-3mm]
{\sc Foces}~&$R$\,=\,40\,000 \\
\hline\\[-2mm]
            7 &   886 &  39 &$\gamma$\,Peg  & B2\,IV &$\beta$\,Cep &Field   &        &  2.834 & $-$0.226  & 24-09-2005 & 2 &  180 & 530\\[-.2mm]
              &       &     &               &        &             &        &  $\pm$ &  0.015 & $ $0.012  &\\[-.2mm]
            A & 22951 & 1123& $o$\,Per      & B0.5\,V&             &Per\,OB2&        &  4.968 & $-$0.017  & 24-09-2005 & 1 &  500 & 280\\[-.2mm]
              &       &     &               &        &             &        &  $\pm$ &  0.008 & $ $0.012  &\\[-.2mm]
            8 & 29248 & 1463&$\nu\,$Eri     &B2\,III &$\beta$\,Cep & Field  &        &  3.930 & $-$0.210  & 14-10-2005 & 1 &  360 & 390\\[-.2mm]
              &       &     &               &        &             &        &  $\pm$ &  0.023 & $ $0.009  &\\[-.2mm]
            9 & 35299 & 1781&               &B1.5\,V &             &Ori\,OB1\,Ia&    &  5.694 & $-$0.210  & 15-10-2005 & 2 & 2400 & 330\\[-.2mm]
              &       &     &               &        &             &        &  $\pm$ &  0.010 & $ $0.007  &\\[-.2mm]
            B & 35468 &1790 &Bellatrix  &B2\,III &             &Ori\,OB1&        &  1.636 & $-$0.224  & 24-09-2005 & 6 &  180 & 520\\[-.2mm]
              &       &     &               &        &             &        &  $\pm$ &  0.007 & $ $0.008  &\\[-.2mm]
           10 & 35708 & 1810&$o$\,Tau       &B2.5\,IV&             &Cas-Tau &        &  4.875 & $-$0.145  & 14-10-2005 & 1 &  600 & 280\\[-.2mm]
              &       &     &               &        &             &        &  $\pm$ &  0.012 & $ $0.006  &\\[-.2mm]
           11 & 36512 & 1855&$\upsilon$\,Ori&B0\,V   &$\beta$\,Cep &Ori\,OB1\,Ic&    &  4.618 & $-$0.264  & 27-09-2005 & 2 &  720 & 500\\[-.2mm]
              &       &     &               &        &             &        &  $\pm$ &  0.013 & $ $0.007  &\\[-.2mm]
           12 & 36822 & 1876&$\phi^1$\,Ori  &B0\,III &             &Ori\,OB1&        &  4.408 & $-$0.162  & 28-09-2005 & 2 &  480 & 420\\[-.2mm]
              &       &     &               &        &             &        &  $\pm$ &  0.006 & $ $0.013  &\\[-.2mm]
           13 & 36960 & 1887&               &B0.5\,V &             &Ori\,OB1\,Ic&    &  4.785 & $-$0.250  & 28-09-2005 & 1 &  240 & 260\\[-.2mm]
              &       &     &               &        &             &        &  $\pm$ &  0.007 & $ $0.010  &\\[-.2mm]
            C & 37023 & 1896& $\theta^1$\,Ori\,D&B0.5\,Vp&         &Ori\,OB\,1d&     &  6.700 & $ $0.080  & 26-09-2001 & 5 & 3080 & 400\\[-.2mm]
              &       &     &               &        &             &        &  $\pm$ &  \dots & $ $\dots  &\\[-.2mm]
           14 &205021 & 8238&$\beta$\,Cep   &B1\,IV  &$\beta$\,Cep &Field   &        &  3.233 & $-$0.222  & 26-09-2005 & 1 &  100 & 310\\[-.2mm]
              &       &     &               &        &             &        &  $\pm$ &  0.014 & $ $0.006  &\\[-.2mm]
           15 &209008 & 8385&18\,Peg        &B3\,III &             &   Field&        &  5.995 & $-$0.120  & 08-10-2005 & 2 & 1800 & 410\\[-.2mm]
              &       &     &               &        &             &        &  $\pm$ &  0.008 & $ $0.014  &\\[-.2mm]
           16 &216916 & 8725&EN\,Lac        &B2\,IV  &$\beta$\,Cep &Lac\, OB1&       &  5.587 & $-$0.144  & 26-09-2005 & 1 &  600 & 270\\[-.2mm]
              &       &     &               &        &             &        &  $\pm$ &  0.015 & $ $0.008  &\\
\hline\\[-3mm]
{\sc Elodie}~&$R$\,=\,42\,000\\
\hline\\[-2mm]
           17 &  3360 &  153&$\zeta$\,Cas   & B2\,IV & SPB         &Cas-Tau &        &  3.661 & $-$0.196  & 14-08-2003 & 4 & 1260 & 310\\[-.2mm]
              &       &     &               &        &             &        &  $\pm$ &  0.017 & $ $0.006  &\\[-.2mm]
            D & 11415 &  542&$\epsilon$\,Cas& B3\,III& Be?         &Cas-Tau &        &  3.370 & $-$0.155  & 12-01-2003 & 2 &  300 & 310\\[-.2mm]
              &       &     &               &        &             &        &  $\pm$ &  0.009 & $ $0.007  &\\[-.2mm]
           18 & 16582 &  779&$\delta$\,Cet  &B2\,IV  &$\beta$\,Cep &Cas-Tau &        &  4.067 & $-$0.219  & 13-01-2003 & 5 & 1250 & 310\\[-.2mm]
              &       &     &               &        &             &        &  $\pm$ &  0.007 & $ $0.008  &\\[-.2mm]
           19 & 34816 & 1756&$\lambda$\,Lep &B0.5\,IV&             &Ori\,OB1&        &  4.286 & $-$0.273  & 25-12-1996 & 1 & 1200 & 250\\[-.2mm]
              &       &     &               &        &             &        &  $\pm$ &  0.005 & $ $0.015  &\\[-.2mm]
           20 &160762 & 6588&$\iota$\,Her   &B3\,IV  & SPB         &Field   &        &  3.800 & $-$0.179  & 28-05-2003 & 2 &  600 & 390\\[-.2mm]
              &       &     &               &        &             &        &  $\pm$ &  0.000 & $ $0.003  &\\[-.2mm]
            E &163472 & 6684&V2052\,Oph     &B2\,IV-V&$\beta$\,Cep &Field   &        &  5.823 & $ $0.089  & 19-08-2003 & 3 & 6000 & 290\\[-.2mm]
              &       &     &               &        &             &        &  $\pm$ &  0.015 & $ $0.003  &\\[-.2mm]
            F &214993 & 8640&12\,Lac        &B2\,III &$\beta$\,Cep &Lac\,OB1&        &  5.253 & $-$0.137  & 10-01-2004 & 1 & 2100 & 250\\[-.2mm]
              &       &     &               &        &             &        &  $\pm$ &  0.018 & $ $0.008  &\\[-.2mm]
            G &218376 & 8797&1\,Cas         &B0.5\,IV&             & Field  &        &  4.850 & $-$0.028  & 12-11-2004 & 1 & 2100 & 290\\[-.2mm]
              &       &     &               &        &             &        &  $\pm$ &  0.009 & $ $0.011  &\\[-.2mm]
\hline\\[-6mm]
 \end{tabular}
\tablefoot{
\tablefoottext{a}{\citet{mer91}; \citet{mor78} for HD\,36960 and HD\,37023.}
\tablefoottext{b}{\citet{hj82}.}
}
\end{table*}

\section{The star sample}\label{sect_obs}
\subsection{Primary target selection}\label{target_selection}
An investigation of the present-day chemical composition of the 
solar neighbourhood requires a careful selection of the star sample 
to minimise observationally-induced systematic bias. 
We took advantage of previous studies of early B-type stars 
\citep[][]{grigsby92,gies92,kilian92,cunha94,andrievsky99,adelman02,lyubimkov04,m07} 
to compile our target list. Criteria were the stars to be\\
{\sc i)} {\em bright}: early B-type stars near the main sequence of magnitude 
$V$\,$<$\,6\,mag are located at distances $<$\,500\,pc, with
such a brightness limit facilitating high-quality spectra to be
obtained with relative ease,\\
{\sc ii)} {\em sharp lined}: low (projected) rotational velocities 
$v \sin i$\,$\lesssim$ 40\,km\,s$^{-1}$ allow spectral line analyses to be 
done at highest
precision, maximising the chances to identify line blends and to place
the continuum unambiguously,\\ 
{\sc iii)} {\em single}: this is to prevent systematic errors in the
analysis as second light from a companion distorts the ratio of
line- to continuum-fluxes; objects in SB1 systems with much fainter
companions or individual components in a visual binary can therefore
still qualify as targets, and\\
{\sc iv)} {\em chemically inconspicuous}: the CP phenomenon is rare among early
B-stars \citep[e.g.][]{Smith96} but spectroscopically classified He-weak or He-strong stars need
to be excluded as segregation processes in their atmospheres have rendered 
them useless to trace the pristine chemical composition.\\ 
In consequence, many of the resulting targets are among the best-studied 
early B-type stars, with multiple analyses reported in the literature.
The sample could therefore be viewed as `ideal' for the proposed purpose.

The distribution of the sample stars in the solar neighbourhood is
visualised in Fig.~\ref{distribution} (anticipating distance
determinations in Sect.~\ref{params}). The stars delineate 
Gould's Belt both in the Galactic plane projection as well as in the
vertical cut through the Galactic disk, with few outliers. They
are highly concentrated towards the disk plane, with few objects
located beyond a vertical distance $z$\,$>$\,100\,pc. 
Members of the two most prominent star-forming regions in the solar 
vicinity are included in the sample, object C is one of the Orion
Trapezium stars, objects 5 and 6 are members of the Sco-Cen association.
The asymmetry of
the star distribution is because our main observing program utilised
telescopes on the northern hemisphere (the celestial equator is roughly
spanned by the connecting line between objects~E~and~9).

\begin{table}[t]
\centering
 \setlength{\tabcolsep}{.18cm}
 \caption[]{IUE spectra used in this study.\\[-6mm] \label{IUE}}
  \begin{tabular}{rlllll}
   \noalign{}
   \hline\hline
   \# & HD     &  SW      & Date       & LW     & Date\\
   \hline\\[-3mm]
   3  & 63922  &  P09511  & 13-07-1980 & R08237 & 13-07-1980\\
   5  & 122980 &  P46857  & 30-01-1993 & P24819 & 30-01-1993\\
   6  & 149438 &  P33008  & 01-03-1988 & P12766 & 01-03-1988\\
   7  & 886    &  P43467  & 25-12-1991 & P22073 & 25-12-1991\\
   8  & 29248  &  P37958  & 06-01-1990 & \ldots & \ldots\\
   9  & 35299  & P18005\tablefootmark{a}& 18-09-1982 & \ldots & \ldots\\
   11 & 36512  &  P08164  & 04-03-1980 & R07097 & 04-03-1980\\
   12 & 36822  & P08595\tablefootmark{a}& 29-03-1980 & R07338\tablefootmark{a} & 29-03-1980\\
   13 & 36960  &  P30541  & 16-03-1987 & P10338 & 16-03-1987\\
   14 & 205021 & P40477\tablefootmark{a}& 28-12-1990 & P19491\tablefootmark{a} & 28-12-1990\\
   15 & 209008 & P20593\tablefootmark{a}& 03-08-1983 & R16508\tablefootmark{a} & 03-08-1983\\
   17 & 3360   &  P26535  & 03-08-1985 & P09140 & 21-09-1986\\
   18 & 16582  &  P29814  & 05-12-1986 & P09634 & 05-12-1986\\
   19 & 34816  &  P08166  & 04-03-1980 & R17279 & 08-03-1984\\
   20 & 160762 &  P42454  & 13-09-1991 & P21228 & 13-09-1991\\
   \hline\\[-7mm]
   \end{tabular}
   \tablefoot{
   \tablefoottext{a}{High-resolution spectrum.}
   }
   \end{table}

\subsection{Observations and data reduction\label{obsdata}}

High-resolution \'echelle spectra at very high signal-to-noise ($S/N$) ratio 
-- ranging from 250 up to over 800 in $B$ -- 
and wide wavelength coverage for 27 objects were obtained, either by 
own observations, or from archives. A summary of the star sample 
and observational details are
given in Table~\ref{observations}. Object identifications include our own
numbering scheme (to facilitate easy identification in the figures and 
other tables),
HD and HR numbers and names of the stars. Spectral types are given and
an indication of variability, basically divided in $\beta$\,Cephei types 
and slowly-pulsating B-stars (SPB)\footnote{Pulsation periods of 
$\beta$\,Cephei stars vary between $\sim$3 to 8\,hours, while those in 
SPB stars vary between $\sim$0.5 to 3\,days.}. 
Membership to one of the nearby OB
associations or to the field population is indicated. Photometric
information covers observed Johnson $V$-magnitudes and colors $B-V$.
Observational details like the date of observation, the number of individual 
exposures per object $N$, the total exposure time and the resulting 
$S/N$-ratio in the $B$ band of the coadded spectrum are also indicated. 
Our philosophy for coaddition of data aims at obtaining a 'snapshot' of 
the stellar spectrum, even for variable stars. Consequently, only exposures 
taken during a small fraction of the same night were coadded, with the one 
exception of a non-variable star.
The analysis will thus yield meaningful time-invariant quantities like 
elemental abundances or fundamental stellar parameters even for variable 
stars, while atmospheric parameters 
like effective temperature and surface gravity 
will be valid only for the moment of observation.
The observational material divides into three subsets.

The first subset includes stars 1--6, for which spectra were obtained
with {\sc Feros} \citep[Fiberfed Extended Range Optical Spectro\-graph,][]{kaufer99} 
on the ESO 2.2m telescope in La Silla/Chile. {\sc Feros} provides a resolving
power of $R$\,$=$\,$\lambda/\Delta\lambda$\,$\approx$\,48\,000, with
2.2 pixels covering a $\Delta\lambda$ resolution element. Of
the entire {\sc Feros} wavelength range, only the part
between $\sim$3800 and 8000\,{\AA} meets our quality criteria for
detailed analysis. The stars have already been subject
to investigation in \citet[NP06]{np06}, 
NP07, NP08 and PNB08. Nevertheless, they are included here as many
details of the analysis were unreported previously.
We refer the reader to NP07 for details of the observations and
the data reduction. 

\begin{figure*}[t!]
\sidecaption
\includegraphics[width=12cm]{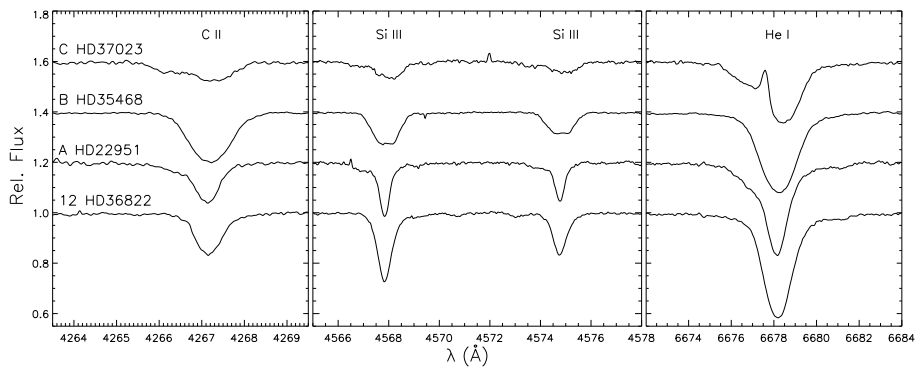}
\caption{Examples of prominent spectral lines of the newly identified
candidate
SB2 systems in our sample, HD\,22951 (B0.5\,V), HD\,35468 (B2\,III) and 
HD\,37023 (B0.5\,V). For comparison the corresponding spectral lines
in the single star HD\,36822 (B0\,III) are shown as well. The emission
peak in \ion{He}{i}\,$\lambda$6678 of the Trapezium star HD\,37023 in
Orion is of nebular origin.
}
\label{sb2}
\end{figure*}

Thirteen stars (7--16, A--C) compose the second subset, which were observed 
with {\sc Foces}
\citep[Fibre Optics Cassegrain \'Ech\-elle Spectrograph,][]{pfeiffer98}
on the 2.2m telescope at Calar Alto/Spain. With the instrument configuration 
chosen a 2-pixel resolution of $R$\,$\approx$\,40\,000 was obtained,
providing wavelength coverage from $\sim$3900 to 9500\,{\AA}. Data
reduction was performed using the {\sc Foces} 
data reduction software \citep{pfeiffer98}, comprising the usual steps of 
bad pixel and cosmic correction, bias and dark current subtraction, 
removal of scattered light, optimum order extraction, flatfielding, 
wavelength calibration using Th-Ar exposures, rectification, and
merging of the \'echelle orders. A major advantage related to the {\sc
Foces} design is that the order tilt for normalisation is much more
homogeneous than in similar spectrographs. This facilitates a
more robust continuum rectification than feasible usually, 
even in the case of broad features like the hydrogen
Balmer lines, which can span more than one
\'echelle order \citep[see][]{Korn02}. 
Finally, multiple exposures of individual stars were
coadded and the spectra were shifted to rest wavelength by
cross-correlation with appropriate synthetic spectra.

Spectra for the third subset of stars (eight stars: 17--20, D--G) were extracted
from the {\sc Elodie} archive \citep{moultaka04}. {\sc Elodie} was an
\'echelle
spectrograph mounted on the 1.93\,m telescope of the Observatoire de
Haute-Provence/France, covering a wavelength range from $\sim$4000 to
6800\,{\AA} at $R$\,$\approx$\,42\,000 \citep[see][]{baranne96}.
The reduced spectra from the archive needed to be normalised
as a last step of post-processing. We used
a spline function to carefully selected continuum points for this
purpose, in analogy to our procedure for the {\sc Feros} data. 
Unfortunately, a reliable normalisation of broad lines in the blue like
H$\gamma$ and H$\delta$ turned out difficult, such that they
were not considered for the analysis. After normalisation, the
individual spectra of one object were coadded in order to
obtain a combined spectrum~of~improved~$S/N$.

In addition to the spectra, which constitute the principal data
for the analysis, various (spectro-)photometric data were adopted
from the literature for constructing spectral energy distributions
(SEDs).
Main sources of Johnson $UBVRIJHK$-photometry were \citet{mer91}, 
\citet{mor78} and \citet{ducati02}, and some $JHK$-magnitudes (only
those with a quality marker `A') were
adopted from the Two Micron All Sky Survey 
\citep[2MASS,][]{Skrutskie06}.
Additionally, flux-calibrated spectra for 15 objects as observed with the 
International Ultraviolet Explorer (IUE) were
extracted from the MAST archive\footnote{\tt http://archive.stsci.edu/}.
Only spectra acquired using a large aperture were considered to avoid
light loss. Preference
was given to low-dispersion data, except for a few cases where only
high-dispersion spectra were available. These data cover the range from
1150 to 1980\,{\AA} for the short (SW) and from 1850 to 3290\,{\AA} for the
long wavelength (LW) range camera. Typically, both wavelength ranges
were observed the
same day. A summary of the individual spectra used in the present work
(data ID number and observation date) is given in Table~\ref{IUE}.

The observational material comprises one of the highest quality 
for the quantitative study of elemental abundances from early B-type stars 
in the solar neighbourhood so far. Obser\-vational bias should thus be reduced 
to a minimum. 

\subsection{Stars excluded from the analysis\label{cleaning}}
Once the spectra were reduced we undertook visual
inspections of the data in the course of the quality as\-sess\-ment.
Four objects were immediately found conspicuous, while three more were found to
be problematic in the analysis. About a quarter of the carefully 
pre-selected sample turned out as candidate binaries or as chemically 
peculiar. In the following we want to briefly address the reasons why 
these cases (stars A--G in Table~\ref{observations}) needed 
to be removed from our sample.\\[-9mm]
\paragraph{A:~\object{HD22951}} The object is a so far unrecognised
candidate SB2 system. All the stronger sharp lines in the spectrum show
weaker and broader absorption dips as well, shifted blueward from the
primary's features. Examples are given in Fig.~\ref{sb2}, where the
corresponding lines in one of the single stars of our sample are also
shown for comparison. The secondary of the system is a cooler 
main-sequence star, of spectral type about B2. All the signatures
can be viewed as undetectable at the resolution of classification spectra.
The secondary's light renders the spec\-trum unsuited for our analysis
technique and the star is therefore excluded from further analysis.\\[-9mm]
\paragraph{B:~\object{HD35468} (Bellatrix)} The star was considered a reference for
the definition of spectral type B2\,III by \citet{walborn71}. However, our
spectrum shows that $\gamma$ Ori is a
candidate SB2 system composed of two very similar components, see
Fig.~\ref{sb2}. This resolves the apparent overluminosity of the star
found by \citet{schroeder04} from an evaluation of its {\sc Hipparcos}
parallax.\\[-9mm]
\paragraph{C:~\object{HD37023}} The case resembles that of HD\,22951, see
Fig.~\ref{sb2}. The companion of HD\,37023 is of slightly earlier type than 
that of HD\,22951, as its contribution to the spectrum is stronger. Note
that while HD\,37023 is a known spectroscopic binary \citep[see][for a
discussion]{Vitrichenko02} the direct detection of a companion in the  
spectrum\footnote{The presence of light from a companion in the
spectrum of HD\,37023 was indicated recently also by 
\citet[his footnote 3]{sergio10}, however without giving further details.} 
offers the possibility to put much tighter constraints on the system, such
that follow-up observations are recommended, like for the two previous
cases.\\[-9mm]
\paragraph{D:~\object{HD11415}} A preliminary analysis of the star yielded
several inconsistencies, which either relate to unrecognised problems
with the {\sc Elodie} spectrum (in particular to the continuum
normalisation) or to some peculiarity of the star. A closer inspection
of the literature revealed that it is listed in the catalogue of Be-stars 
of \citet{jaschek82}. All Balmer lines and in particular H$\alpha$ are in 
absorption in the available spectrum, but we cannot rule out low-level
emission filling in the inner line wings (imitating normalisation
problems). Moreover, \citet{lanzafame95} describe 
HD\,11415 as CP star of the He-weak type. These factors prompted us to exclude
HD\,11415 from the detailed analysis.\\[-9mm]
\paragraph{E:~\object{HD163472}} A preliminary analysis of the star
indicated inconsistencies of the solution, which were unexpected in
view of the `normal' appearance implied by the analysis of
\citet{m07}. However, a detailed study by \citet{neiner03} found
HD\,163472 to be chemically peculiar, in particular it is a He-strong star.
Diffusion in the magnetic atmosphere renders this otherwise highly interesting
star useless for our purpose.\\[-9mm]
\paragraph{F:~\object{HD214993}} The star is one of the most intensely
studied $\beta$\,Cephei pulsators \citep[and references therein]{Desmet09}.
Our preliminary modelling encounters difficulties, which may be
resolved assuming helium-enrichment of the atmosphere. This requires
further investigation and in consequence we exclude the star from the
present work program.\\[-9mm]
\paragraph{G:~\object{HD218376}} The star shows conspicuously broad
lines. Whether this may be interpreted as unresolved binarity --
requiring two about similar components -- or whether a different
explanation needs to be found has to be decided by further investigations.

\section{Spectrum synthesis in non-LTE}\label{sect_the}

The non-LTE line-formation computations follow the
methodology discussed in detail in our previous studies for H and He
in NP07, for C in NP06 and NP08, and for N, Ne, Mg, Si and Fe in PNB08.
In brief, a non-LTE approach is employed to solve
the restricted non-LTE problem on the basis of prescribed LTE atmos\-pheres.
This technique provides an efficient way to compute realistic synthetic
spectra in all cases where the atmos\-pheric structure is close to LTE, 
like for the early B-type main sequence stars analysed here (see NP07).
The computational efforts can thus be focussed on robust
non-LTE line-formation calculations. The validity of the approach was
recently verified by direct comparison with hydrodynamic line-blanketed 
non-LTE model atmospheres (NS11). This approach has been
equally successful in improving model fits beyond the field of 
massive B-type stars, e.g. in low-mass subdwarf B-stars \citep{p06b}
and B-type extreme helium stars \citep{pbhj05,p06c}.

\subsection {Models and programs\label{models}}

The model atmos\-pheres were computed with the {\sc Atlas9}~code \citep{kur93b}
which assumes plane-parallel geometry,~chemical homogeneity, as well as 
hydrostatic, radiative  and local thermodynamic equilibrium (LTE). 
Line blanketing was realised here by means of opacity distribution functions
\citep[ODFs,][]{kur93a}. Solar abundances of \citet{gs98} were adopted in all
computations. The model atmospheres were held fixed in the non-LTE calculations.
Non-LTE level populations and model spectra were obtained with
recent versions of {\sc Detail} and {\sc Surface}
\citep[both updated by K.\,Butler]{gid81,but_gid85}. The 
coupled~radiative transfer 
and statistical equilibrium equations were solved with {\sc Detail}, 
employing an accelerated lambda itera\-tion scheme of \citet{rh91}. 
This allowed even complex ions to be treated in a
realistic way. Synthetic spectra were calculated with {\sc Surface}, using refined
line-broadening theories. 
Continuous opacities due to hydrogen and helium were considered in non-LTE and
line blocking was accounted for in LTE via Kurucz' ODFs.
Microturbulence was considered in a consistent way throughout all
computation steps: in the selection of appropriate ODFs for realising
line blanketing and line blocking in the atmospheric structure and
non-LTE level populations determination, and for the formal solution. 

\begin{table}[t!]
\footnotesize
\caption[]{Model atoms for non-LTE calculations.\\[-6mm] \label{atoms}}
\setlength{\tabcolsep}{.15cm}
\begin{tabular}{ll}
\hline
\hline
\footnotesize
            Ion     &  Model atom \\
\hline\\[-3mm] 
     H          &  \citet{pb04}\\
\ion{He}{i/ii}  &  \citet{p05}\\
\ion{C}{ii-iv}  &  NP06, NP08\\
\ion{N}{ii}     &  \citet{pb01}, updated\tablefootmark{a}\\
\ion{O}{i/ii}   &  \citet{p00}, \citet{bb88}, updated\tablefootmark{a}\\
\ion{Ne}{i/ii}  &  \citet{mb08}, updated\tablefootmark{a}\\
\ion{Mg}{ii}    &  \citet{p01}\\
\ion{Si}{iii/iv} &  \citet{bb90}\\
\ion{Fe}{ii/iii}&  \citet{b98}, \citet{m07}, corrected\tablefootmark{b}\\
            \hline\\[-5mm]
           \end{tabular}
\tablefoot{
\tablefootmark{a}{See Table~7 for details.}
\tablefootmark{b}{See Sect.~\ref{models}.}}
\end{table}

Non-LTE level populations and the synthetic spectra of all elements
were computed using our most recent model atoms listed in Table~\ref{atoms}. 
Updates of some of the published models were carried out introducing improved 
oscillator strengths and collisional data. These models were previously tested 
in NP08 and PNB08 for early B-type stars covering a wide parameter range.
A problem with the line-formation calculations for \ion{Fe}{iii} was
identified in the course of the present work. Several 
high-lying energy levels were previously erroneously treated in LTE in the formal 
solution with {\sc Surface}, despite correct non-LTE level populations were
provided by {\sc Detail}.  Higher 
iron abundances for the hotter objects (e.g. by
0.16\,dex for $\tau$\,Sco and $\lesssim$0.05\,dex for the majority of
stars) result after implementing the corrected Fe model atom, 
removing a slight artificial trend 
with temperature found previously.

\subsection {Grids of synthetic spectra}\label{grids}
For the present work a set of grids of synthetic
spectra was computed with {\sc Atlas9}, {\sc Detail} and {\sc Surface}
following the same procedure as in our previous papers.
Large independent grids of H/He, C, N, O, Mg and Si encompass effective
temperatures from 15\,000 to
35\,000\,K in steps of 1000\,K, surface gravities $\log g$ from 3.0 to
4.5 (cgs units) in steps of 0.1\,dex, microturbulences from 0 to 12\,km\,s$^{-1}$
in steps of 2\,km\,s$^{-1}$ and metal abundances within 1\,dex
centered on the B-star abundance values of PNB08 in steps of 0.1\,dex.
Hydrogen and helium abundances are set to the values derived by PNB08.
The lower limit of the surface gravity for each value of 
temperature/microturbulence is constrained by the convergence of {\sc Atlas9}.
All grids have been successfully tested by reproducing results from PNB08.
For Ne and Fe (the computationally most demanding species) microgrids -- 
varying abundance only -- were
computed per star once all stellar parameters were determined with the
larger pre-computed grids. 

\begin{figure}[!t]
\hspace{.8cm}
\includegraphics[width=1.\linewidth]{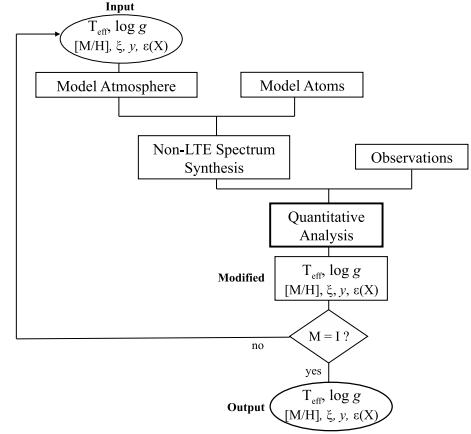}
\vspace{-8mm}
\caption{Schematic diagram of our analysis method. 
See Sect.~\ref{semi-automatic} for details.}
\label{diagram}
\end{figure}

\section{Spectral analysis}\label{sect_an}
Our analysis method is based on the simultaneous reproduction of all spectroscopic
indicators (Sect.~\ref{indic}) via an iterative line-fitting procedure 
aiming to derive atmospheric parameters and chemical abundances self-consistently. 
In contrast to common strategies in stellar spectroscopy, this analysis technique
takes full advantage of the information encoded in the {\em line
profiles} at different
wavelength ranges simultaneously. Integrated quantities like equivalent widths
W$_{\lambda}$ are not measured in this approach. 
The stellar parameters primarily derived here are 
the effective temperature $T_\mathrm{eff}$, surface gravity $\log g$, 
microturbulence $\xi$, (radial-tangential) 
macroturbulence $\zeta$, projected
rotational velocity $v\,\sin\,i$ and elemental abundances 
$\varepsilon(X)\,=\,\log$($X$/H)$\,+\,12$.

\subsection{Semi-automation of the analysis
procedure\label{semi-automatic}}

The basic analysis methodology was introduced and applied
to a strategically chosen star sample covering a broad parameter
range in \citet{n07}, NP07 and NP08. A major complication in that work was that
both the model atom for carbon and the stellar parameters had to be 
constrained simultaneously. 
Thus, one global problem needed to be solved -- to select among the available
atomic data the one combination that facilitated reproduction of the
observed lines throughout the entire sample equally well -- and many
individual problems -- the determination of atmospheric models that
described the sample stars best. The only way to realise this was via
the computation of numerous small grids per star for different sets of stellar parameters 
and atomic data, and interactive work to decide how to improve based on
the quality of match between the models and observation. Over 100 interactive 
iterations were needed to find a self-consistent solution, i.e. selection of the
best atomic data and determination of stellar parameters and carbon
abundances for six stars covering a broad temperature range.
The benefit from this time-consuming approach 
was twofold: first, a highly robust model atom was provided for further
applications, and second, copious experience was gained in the
identification of sources of systematic error that helps to minimise 
such uncertainties
in work thereafter. This lead, e.g. to updates of model atoms for other
elements, see PNB08 and Table~\ref{atoms}. All in all, a solid basis 
for highly accurate (reduction of systematic uncertainty to the greatest possible 
extent) and precise (low statistical uncertainties) analyses was thus laid.

Once the global problem of the model atom construction and testing is
solved, one is faced with the much easier task that only stellar
parameters and elemental abundances need to be constrained. The basic
analysis methodology is outlined in Fig.~\ref{diagram}. The individual
steps could be realised in form of small grid computations like
outlined before, but this is too inefficient for investigating
larger star samples. 
We have therefore replaced the calculation of dedicated small grids of
synthetic spectra (comprising the boxes `Model Atmosphere', `Model
Atom' and `Non-LTE Spectrum Synthesis' in Fig.~\ref{diagram}) by a
pre-computed comprehensive grid of models (in total of the order
$\sim$100\,000 synthetic spectra), as described in Sect.~\ref{grids}.
The second new ingredient for the present work was the adaptation of 
a powerful fitting routine for the semi-automatic comparison of
observed and theoretical spectra (the box
`Quantitative Analysis' in Fig.~\ref{diagram}).
{\sc Spas}\footnote{Spectrum Plotting and Analysing Suite, {\sc Spas} 
\citep{Hirsch09}.} provides the means to interpolate between model
grid points for up to three parameters simultaneously and allows to 
apply instrumental, rotational and (radial-tangential) macrobroadening functions
to the resulting theoretical profiles. The program uses the downhill
simplex algorithm \citep{NeMe65} to minimise $\chi^2$ in order to 
find a good fit to the observed spectrum. 

\begin{table}[t!]
\centering
\caption[]{
Spectroscopic indicators for for $T_\mathrm{eff}$ and $\log g$
determination.\\[-6mm]\label{indicators}}
 \setlength{\tabcolsep}{.045cm}
 \tiny
 \begin{tabular}{rrrcc@{\hspace{.1mm}}cc@{\hspace{.1mm}}c@{\hspace{.1mm}}cc@{\hspace{.1mm}}cc@{\hspace{.1mm}}cc@{\hspace{.1mm}}cc@{\hspace{.1mm}}c}
 \noalign{}
\hline
\hline
\#&HD&$T_\mathrm{eff}$& H & \ion{He}{i} & \ion{He}{ii}                 &
\ion{C}{ii} & \ion{C}{iii}  & \ion{C}{iv} & \ion{O}{i} & \ion{O}{ii} &
\ion{Ne}{i} &  \ion{Ne}{ii} & \ion{Si}{iii}& \ion{Si}{iv} & \ion{Fe}{ii}& \ion{Fe}{iii}\\
 &  & 10$^3$\,K\\[-.1mm]
\hline\\[-2mm]
11&36512 & 33.4   & $\bullet$ & \multicolumn{2}{c}{\fbox{$\bullet$~~~~~$\bullet$}}& \multicolumn{3}{c}{\fbox{\,$\bullet$~~~~~$\bullet$~~~~~$\bullet$\,}} &                                        & $\bullet$ &           &                             $\bullet$  &\multicolumn{2}{c}{\fbox{$\bullet$~~~~~$\bullet$}}& & $\bullet$\\[.5mm]
6 &149438& 32.0   & $\bullet$ & \multicolumn{2}{c}{\fbox{$\bullet$~~~~~$\bullet$}}& \multicolumn{3}{c}{\fbox{\,$\bullet$~~~~~$\bullet$~~~~~$\bullet$\,}} &                                        & $\bullet$ & \multicolumn{2}{c}{\fbox{$\bullet$~~~~~$\bullet$}} &\multicolumn{2}{c}{\fbox{$\bullet$~~~~~$\bullet$}}& & $\bullet$\\[.5mm]
3 &63922 & 31.2   & $\bullet$ & \multicolumn{2}{c}{\fbox{$\bullet$~~~~~$\bullet$}}& \multicolumn{3}{c}{\fbox{\,$\bullet$~~~~~$\bullet$~~~~~$\bullet$\,}} &                                        & $\bullet$ &           &                             $\bullet$  &\multicolumn{2}{c}{\fbox{$\bullet$~~~~~$\bullet$}}& & $\bullet$\\[.5mm]
19&34816 & 30.4   & $\bullet$ & \multicolumn{2}{c}{\fbox{$\bullet$~~~~~$\bullet$}}& \multicolumn{2}{c}{\fbox{$\bullet$~~~~~$\bullet$}} &                 &                                        & $\bullet$ &  \multicolumn{2}{c}{\fbox{$\bullet$~~~~~$\bullet$}}&\multicolumn{2}{c}{\fbox{$\bullet$~~~~~$\bullet$}}& & $\bullet$\\[.5mm]
12&36822 & 30.0   & $\bullet$ & \multicolumn{2}{c}{\fbox{$\bullet$~~~~~$\bullet$}}& \multicolumn{3}{c}{\fbox{\,$\bullet$~~~~~$\bullet$~~~~~$\bullet$\,}} &                                        & $\bullet$ & \multicolumn{2}{c}{\fbox{$\bullet$~~~~~$\bullet$}} &\multicolumn{2}{c}{\fbox{$\bullet$~~~~~$\bullet$}}& & $\bullet$\\[.5mm]
13&36960 & 29.0   & $\bullet$ & \multicolumn{2}{c}{\fbox{$\bullet$~~~~~$\bullet$}}& \multicolumn{2}{c}{\fbox{$\bullet$~~~~~$\bullet$}} &                 &                                        & $\bullet$ & \multicolumn{2}{c}{\fbox{$\bullet$~~~~~$\bullet$}} &\multicolumn{2}{c}{\fbox{$\bullet$~~~~~$\bullet$}}& & $\bullet$\\[.5mm] 
1 &36591 & 27.0   & $\bullet$ & \multicolumn{2}{c}{\fbox{$\bullet$~~~~~$\bullet$}}& \multicolumn{2}{c}{\fbox{$\bullet$~~~~~$\bullet$}} &                 &                                        & $\bullet$ & \multicolumn{2}{c}{\fbox{$\bullet$~~~~~$\bullet$}} &\multicolumn{2}{c}{\fbox{$\bullet$~~~~~$\bullet$}}& & $\bullet$\\[.5mm]
14&205021& 27.0   & $\bullet$ & \multicolumn{2}{c}{\fbox{$\bullet$~~~~~$\bullet$}}& \multicolumn{2}{c}{\fbox{$\bullet$~~~~~$\bullet$}} &                 &                                        & $\bullet$ & \multicolumn{2}{c}{\fbox{$\bullet$~~~~~$\bullet$}} &\multicolumn{2}{c}{\fbox{$\bullet$~~~~~$\bullet$}}& & $\bullet$\\[.5mm] 
2 &61068 & 26.3   & $\bullet$ & \multicolumn{2}{c}{\fbox{$\bullet$~~~~~$\bullet$}}& \multicolumn{2}{c}{\fbox{$\bullet$~~~~~$\bullet$}} &                 &                                        & $\bullet$ & \multicolumn{2}{c}{\fbox{$\bullet$~~~~~$\bullet$}} &\multicolumn{2}{c}{\fbox{$\bullet$~~~~~$\bullet$}}& & $\bullet$\\[.5mm]
9 &35299 & 23.5   & $\bullet$ & $\bullet$ &                                       & \multicolumn{2}{c}{\fbox{$\bullet$~~~~~$\bullet$}} &                 & \multicolumn{2}{c}{\fbox{$\bullet$~~~~~$\bullet$}} & \multicolumn{2}{c}{\fbox{$\bullet$~~~~~$\bullet$}} &\multicolumn{2}{c}{\fbox{$\bullet$~~~~~$\bullet$}}& \multicolumn{2}{c}{\fbox{$\bullet$~~~~~$\bullet$}}\\[.5mm]  
16&216916& 23.0   & $\bullet$ & $\bullet$ &                                       & \multicolumn{2}{c}{\fbox{$\bullet$~~~~~$\bullet$}} &                 & \multicolumn{2}{c}{\fbox{$\bullet$~~~~~$\bullet$}} & \multicolumn{2}{c}{\fbox{$\bullet$~~~~~$\bullet$}} &\multicolumn{2}{c}{\fbox{$\bullet$~~~~~$\bullet$}}& \multicolumn{2}{c}{\fbox{$\bullet$~~~~~$\bullet$}}\\[.5mm]   
4 &74575 & 22.9   & $\bullet$ & $\bullet$ &                                       & \multicolumn{2}{c}{\fbox{$\bullet$~~~~~$\bullet$}} &                 & \multicolumn{2}{c}{\fbox{$\bullet$~~~~~$\bullet$}} & \multicolumn{2}{c}{\fbox{$\bullet$~~~~~$\bullet$}} &\multicolumn{2}{c}{\fbox{$\bullet$~~~~~$\bullet$}}& \multicolumn{2}{c}{\fbox{$\bullet$~~~~~$\bullet$}}\\[.5mm]
7 &886   & 22.0   & $\bullet$ & $\bullet$ &                                       & \multicolumn{2}{c}{\fbox{$\bullet$~~~~~$\bullet$}} &                 & \multicolumn{2}{c}{\fbox{$\bullet$~~~~~$\bullet$}} & $\bullet$ &                                        &\multicolumn{2}{c}{\fbox{$\bullet$~~~~~$\bullet$}}& \multicolumn{2}{c}{\fbox{$\bullet$~~~~~$\bullet$}}\\[.5mm]
8 &29248 & 22.0   & $\bullet$ & $\bullet$ &                                       & \multicolumn{2}{c}{\fbox{$\bullet$~~~~~$\bullet$}} &                 & \multicolumn{2}{c}{\fbox{$\bullet$~~~~~$\bullet$}} & $\bullet$ &                                        &\multicolumn{2}{c}{\fbox{$\bullet$~~~~~$\bullet$}}& \multicolumn{2}{c}{\fbox{$\bullet$~~~~~$\bullet$}}\\[.5mm] 
18&16582 & 21.3   & $\bullet$ & $\bullet$ &                                       & \multicolumn{2}{c}{\fbox{$\bullet$~~~~~$\bullet$}} &                 & \multicolumn{2}{c}{\fbox{$\bullet$~~~~~$\bullet$}} & $\bullet$ &                                        &\multicolumn{2}{c}{\fbox{$\bullet$~~~~~$\bullet$}}& \multicolumn{2}{c}{\fbox{$\bullet$~~~~~$\bullet$}} \\[.5mm]
5 &122980& 20.8   & $\bullet$ & $\bullet$ &                                       & \multicolumn{2}{c}{\fbox{$\bullet$~~~~~$\bullet$}} &                 & \multicolumn{2}{c}{\fbox{$\bullet$~~~~~$\bullet$}} & $\bullet$ &                                        &$\bullet$&& \multicolumn{2}{c}{\fbox{$\bullet$~~~~~$\bullet$}}\\[.5mm]
10&35708 & 20.7   & $\bullet$ & $\bullet$ &                                       & \multicolumn{2}{c}{\fbox{$\bullet$~~~~~$\bullet$}} &                 & \multicolumn{2}{c}{\fbox{$\bullet$~~~~~$\bullet$}} & $\bullet$ &                                        &\multicolumn{2}{c}{\fbox{$\bullet$~~~~~$\bullet$}}& \multicolumn{2}{c}{\fbox{$\bullet$~~~~~$\bullet$}}\\[.5mm] 
17&  3360& 20.7   & $\bullet$ & $\bullet$ &                                       & \multicolumn{2}{c}{\fbox{$\bullet$~~~~~$\bullet$}} &                 & \multicolumn{2}{c}{\fbox{$\bullet$~~~~~$\bullet$}} & $\bullet$ &                                        &\multicolumn{2}{c}{\fbox{$\bullet$~~~~~$\bullet$}}& \multicolumn{2}{c}{\fbox{$\bullet$~~~~~$\bullet$}}\\[.5mm]
20&160762& 17.5   & $\bullet$ & $\bullet$ &                                       &  $\bullet$                                       & &                 & \multicolumn{2}{c}{\fbox{$\bullet$~~~~~$\bullet$}} & $\bullet$ &                                        &$\bullet$&& \multicolumn{2}{c}{\fbox{$\bullet$~~~~~$\bullet$}}\\[.5mm]
15&209008& 15.8   & $\bullet$ & $\bullet$ &                                       &  $\bullet$                                       & &                 & \multicolumn{2}{c}{\fbox{$\bullet$~~~~~$\bullet$}} & $\bullet$ &                                        &$\bullet$&& \multicolumn{2}{c}{\fbox{$\bullet$~~~~~$\bullet$}}\\[.5mm] 
\hline
\vspace{-4mm}
\end{tabular}
\tablefoot{The boxes denote ionization equilibria.}
\end{table}

Interactive work in some decisive steps on the analysis with {\sc Spas} 
paid off as much more accurate results could be obtained.
Crucial was the selection of the appropriate spectroscopic indicators 
(Sect.~\ref{indic}) for the parameter determination which may vary
from star to star upon availability of specific spectral lines
(depending on stellar temperature, spectrum quality and the observed 
wavelength coverage). All spectral lines unsuited for analysis because
of e.g. blends, low S/N, uncorrectable normalisation problems, 
incomplete correction of cosmics, or known shortcomings in the
modelling needed to be excluded. Also a verification and, possibly,
correction of the automatic continuum placement lead to a gain in
precision.
Every element was analysed independently (passing through the loop
procedure in Fig.~\ref{diagram}) and some interactive iterations 
for fine-tuning the parameter determination were needed in order to 
find a unique solution that reproduces all indicators simultaneously.
This facilitated also to derive realistic uncertainties for the
stellar parameters, as the formal errors determined by {\sc Spas} 
(via bootstrapping) were unrealistically low. Instead, the standard
deviations around the average parameter values were adopted, as 
derived from the various independent spectral indicators. Likewise,
uncertainties of elemental abundances were determined from the
line-to-line scatter found from the analysis of the individual features.

Finally, it was thus possible to derive a simultaneous, self-consistent 
solution for atmospheric parameters and chemical abundances, and also to
quantify their statistical uncertainties. The novel approach provides
results meeting the same quality standard as our previous work (for
test purposes and consistency checks we therefore included the
previously analysed objects in the present sample, stars 1-6 in the
tables). Its advantages are a higher degree of objectivity than `by
eye' fits and a far higher efficiency, hence allowing larger star samples 
to be analysed.

\subsection{Stellar parameter and abundance determination}\label{indic}
Special emphasis was given to use multiple
indicators in order to minimise the chance of the stellar atmospheric
parameters and chemical abundance determination being biased by residual
systematic errors. The following spectroscopic indicators were utilised 
in the quantitative analysis:

\noindent $\bullet$ $T_\mathrm{eff}$:~~all available H and He  
lines, and multiple {\em independent} ioni\-zation equilibria; confirmation 
via spectral energy distributions (SEDs);\vspace{.5mm}

\noindent $\bullet$ $\log g$:~~wings of all available hydrogen lines and multiple 
ioni\-zation equilibria; confirmation via {\sc Hipparcos} distances; \vspace{.5mm}

\noindent $\bullet$ $\xi$:~~several elements with spectral lines of 
different strength 
enforcing no correlation between $\varepsilon(X)$ and the strength of the lines
(equivalent to $\varepsilon(X)$ being
independent of $W_\lambda$);\vspace{.5mm}

\noindent $\bullet$ $v\sin i$ and $\zeta$:~~metal line profiles;\vspace{.5mm}

\noindent $\bullet$ $\varepsilon(X)$:~~a comprehensive set of metal 
lines.\vspace{.5mm}

The parameter determination started with the analysis of the hydrogen
and helium lines. When a good simultaneous fit to most H and He lines
was achieved -- $T_\mathrm{eff}$ and $\log g$ were then typically constrained 
to better than $\sim$5\% and 0.1--0.2\,dex, respectively, 
for this high-quality set of stellar spectra -- the procedure
commenced to consider lines of other elements. 
Ionization equilibria, i.e.~the requirement that lines
from different ions of an element have to indicate the same chemical
abundance, facilitated a fine-tuning of the previously derived parameters.
The selection of ionization equilibria to be analysed depends
primarily on $T_\mathrm{eff}$. Table~\ref{indicators} summarises the 
spectroscopic indicators employed for the $T_\mathrm{eff}$ and $\log g$ 
determination of each sample star, sorted by temperature. 
Elements that show lines of three ionization stages simultaneously in the
spectrum are most valuable as they can in principle constrain both
$T_\mathrm{eff}$ and $\log g$ at once. Examples are C\,{\sc ii/iii/iv}
or Si\,{\sc ii/iii/iv}\footnote{Note that we analyse Si\,{\sc iii/iv} lines
only because the model atom employed here underestimates silicon abundances
derived from Si\,{\sc ii} lines, see \citet{sergio10} for a discussion.} 
in early B-type stars. When lines from only two ionization stages
of an element are present in the spectra, then more indicators are required
for the parameter determination. Usually in the literature, a few hydrogen 
lines are analysed and one ionization equilibrium is established. On the
other hand, we try to establish typically 4-5 independent ionization equilibria
in addition to fitting all available hydrogen lines, which is unprecedented. 
Finally, the resulting model fluxes were compared with
the observed SEDs. While such a high accuracy in the
$T_\mathrm{eff}$-determination as with our spectroscopic
approach can not be achieved by SED fitting alone, it provides a valuable
consistency check, which can also facilitate the detection of
otherwise unrecognised cooler companions.  

Microturbulence was for a long time an {\em ad-hoc} 
fit parameter that was employed to remove correlations of abundance 
with $W_\lambda$, and often a different $\xi$ was adopted for different elements.
Only recently, a physical explanation for the phenomenon of
microturbulence in hot stars was suggested, 
likely being a consequence of subsurface convection
\citep{cantiello09}, similar to the case of solar-type stars.
Microturbulence needs to be constrained simultaneously along with
$T_\mathrm{eff}$ and $\log g$, such that its determination was a crucial part 
of our iteration procedure. The reason for this is that an 
inappropriately chosen microturbulence may lead to substantial shifts 
in $T_\mathrm{eff}$ from the ionization equilibria analysis, see
Fig.~5 in \citet{np10b}, which may remain unnoticed in the usual
approach of using a minimum set of indicators for the parameter
determination. Consequently, several elements were analysed for
deriving the microturbulent velocity. Our primary indicators were 
the Si, O and C lines, but the results were checked for consistency 
with the lines from the other elements as well.

The detailed analysis of line profiles shows that rotational
broadening alone is often not sufficient to explain the observed line shapes
in hot stars. Agreement can be achieved when introducing a
radial-tangential anisotropic macroturbulence \citep[p.
433f.]{Gray05} as additional broadening agent, see e.g. Fig.~11 of
\citet{p06a}.
Consideration of macroturbulence is therefore essential for meaningful
line-fits using $\chi^2$-minimisation.
A physical explanation of macroturbulence in hot stars was also 
only recently suggested, likely being a collective effect of (non-radial) 
pulsations \citep{aerts09}. 

Usually in stellar analyses, once the stellar parameters are fixed one
commences with the abundance determination, treating this as an
essentially independent step. In our approach the abundance and
stellar parameter determination are tightly related because of the use
of ionization equilibria. In consequence, only few species (those not
appearing as ionization equilibria in Table~\ref{indicators}) are left
to finalise the analysis. Another difference to typical literature
studies is the large number of spectral lines evaluated by us per
species, and the consistency achieved from the different ionization 
stages of the various elements. All the various improvements in 
observations, modelling and analysis methodology facilitated analyses 
at much higher precision and accuracy to be achieved than possible in 
standard works. The quality of the analyses could be
retained over a large parameter space, spanning nearly 20\,000\,K in
$T_\mathrm{eff}$ and ranging from close to the zero-age main sequence
 (ZAMS) to the giant stage. Consequently, an excellent match of the
computed and the observed spectra is achieved globally and in the
details, see the end of Sect.~\ref{params} for a discussion.

\begin{table}[t!]
\centering
\caption[]{Stellar parameters of the program stars\tablefootmark{a}.\\[-6mm]\label{bigtable}}
 \setlength{\tabcolsep}{.065cm}
 \begin{tabular}{rrr@{\hspace{-1.5mm}}rrrrrrrrrr}
 \noalign{}
\hline
\hline
\#&HD & & $T_\mathrm{eff}$ &$\log\,g$ & ~$\xi$ &$v\sin\,i$& $\zeta$& $E(B-V)$ & $V_0$ & $M_\mathrm{ev}$ & $d_\mathrm{spec}$ & $d_\mathrm{HIP}$\\[-.3mm]
\cline{6-8}
&    & & 10$^3$\,K    & (cgs)   &
\multicolumn{3}{c}{{km\,s$^{-1}$}} & mag
&mag & $M_{\sun}$ & pc & pc\\[-.2mm]
\hline\\[-1mm]
1 &36591 &     & 27.0 & 4.12 & 3 & 12  & \dots& 0.06 & 5.16 & 12.3 & 408 & \dots\\[-.3mm]
  &      &$\pm$&  0.3 & 0.05 & 1 &  1  & \dots& 0.00 & 0.01 &  0.3 &  26 & \dots\\[1mm]
2 &61068 &     & 26.3 & 4.15 & 3 & 14  & 20   & 0.07 & 5.49 & 11.6 & 434 & \dots\\[-.3mm]
  &      &$\pm$&  0.3 & 0.05 & 1 &  2  &  1   & 0.01 & 0.03 &  0.3 &  28 & \dots\\[1mm]
3 &63922 &     & 31.2 & 3.95 & 8 & 29  & 37   & 0.09 & 3.82 & 18.9 & 389 & \dots\\[-.3mm]
  &      &$\pm$&  0.3 & 0.05 & 1 &  4  &  8   & 0.01 & 0.02 &  0.5 &  25 & \dots\\[1mm]
4 &74575 &     & 22.9 & 3.60 & 5 & 11  & 20   & 0.05 & 3.54 & 11.7 & 301 & 270  \\[-.3mm]
  &      &$\pm$&  0.3 & 0.05 & 1 &  2  &  1   & 0.00 & 0.01 &  1.2 &  24 &  10  \\[1mm]
5 &122980&     & 20.8 & 4.22 & 3 & 18  & \dots& 0.01 & 4.32 &  7.5 & 150 & 156  \\[-.3mm]
  &      &$\pm$&  0.3 & 0.05 & 1 &  1  & \dots& 0.01 & 0.02 &  0.2 &  10 &   5  \\[1mm]
6 &149438&     & 32.0 & 4.30 & 5 &  4  &  4   & 0.03 & 2.73 & 15.8 & 143 & 145  \\[-.3mm]
  &      &$\pm$&  0.3 & 0.05 & 1 &  1  &  1   & 0.01 & 0.02 &  0.7 &   9 &  11  \\[1mm]
7 &886   &     & 22.0 & 3.95 &  2 &  9 &  8   & 0.00 & 2.83 &  9.2 & 120 & 120  \\[-.3mm]
  &      &$\pm$&  0.4 & 0.05 &  1 &  2 &  2   & 0.01 & 0.04 &  0.3 &   8 &   8  \\[1mm]
8 &29248 &     & 22.0 & 3.85 &  6 & 26 & 15   & 0.01 & 3.90 &  9.5 & 225 & 207  \\[-.3mm]
  &      &$\pm$& 0.25 & 0.05 &  1 &  2 &  5   & 0.01 & 0.04 &  0.3 &  15 &   8  \\[1mm]
9 &35299 &     & 23.5 & 4.20 &  0 &  8 & \dots& 0.02 & 5.64 &  9.2 & 344 & 269  \\[-.3mm]
  &      &$\pm$&  0.3 & 0.05 &  1 &  1 & \dots& 0.01 & 0.02 &  0.3 &  22 &  24  \\[1mm]
10&35708 &     & 20.7 & 4.15 &  2 & 25 & 17   & 0.06 & 4.69 &  7.6 & 192 & 192  \\[-.3mm]
  &      &$\pm$&  0.2 & 0.07 &  1 &  2 &  5   & 0.01 & 0.02 &  0.2 &  17 &   8  \\[1mm]
11&36512 &     & 33.4 & 4.30 &  4 & 20 & 10   & 0.02 & 4.54 & 18.0 & 366 & \dots\\[-.3mm]
  &      &$\pm$&  0.2 & 0.05 &  1 &  2 &  5   & 0.01 & 0.03 &  0.7 &  24 & \dots\\[1mm]
12&36822 &     & 30.0 & 4.05 &  8 & 28 & 18   & 0.11 & 4.06 & 16.2 & 338 & 333  \\[-.3mm]
  &      &$\pm$&  0.3 & 0.10 &  1 &  2 &  5   & 0.01 & 0.04 &  1.1 &  46 &  28  \\[1mm]
13&36960 &     & 29.0 & 4.10 &  4 & 28 & 20   & 0.02 & 4.74 & 14.4 & 392 & \dots\\[-.3mm]
  &      &$\pm$&  0.3 & 0.07 &  1 &  3 &  7   & 0.01 & 0.03 &  0.6 &  36 & \dots\\[1mm]
14&205021&     & 27.0 & 3.95 &  4 & 28 & 20   & 0.03 & 3.13 & 13.3 & 201 & 210  \\[-.3mm]
  &      &$\pm$& 0.45 & 0.05 &  1 &  3 &  7   & 0.01 & 0.03 &  0.5 &  13 &  13  \\[1mm]
15&209008&     & 15.8 & 3.75 &  4 & 15 & 10   & 0.04 & 5.87 &  5.8 & 372 & \dots\\[-.3mm]
  &      &$\pm$&  0.2 & 0.05 &  1 &  3 &  3   & 0.01 & 0.04 &  0.2 &  25 & \dots\\[1mm]
16&216916&     & 23.0 & 3.95 &  0 & 12 & \dots& 0.08 & 5.33 &  9.8 & 405 & \dots\\[-.3mm]
&        &$\pm$&  0.2 & 0.05 &  1 &  1 & \dots& 0.01 & 0.03 &  0.3 &  26 & \dots\\[1mm]
17&  3360&     & 20.75& 3.80 &  2 & 20 & 12   & 0.02 & 3.61 &  8.9 & 191 & 182  \\[-.3mm]
  &      &$\pm$&  0.2 & 0.05 &  1 &  2 &  5   & 0.01 & 0.03 &  0.3 &  12 &   5  \\[1mm]
18& 16582&     & 21.25& 3.80 &  2 & 15 & 10   & 0.00 & 4.07 &  9.3 & 241 & 199  \\[-.3mm]
  &      &$\pm$&  0.4 & 0.05 &  1 &  2 &  5   & 0.01 & 0.03 &  0.3 &  16 &   6  \\[1mm]
19& 34816&     & 30.4 & 4.30 &  4 & 30 & 20   & 0.00 & 4.29 & 14.4 & 264 & 261  \\[-.3mm]
  &      &$\pm$&  0.3 & 0.05 &  1 &  2 &  7   & 0.02 & 0.05 &  0.4 &  18 &  16  \\[1mm]
20&160762&     & 17.5 & 3.80 &  1 &  6 & \dots& 0.00 & 3.80 &  6.7 & 157 & 139  \\[-.3mm]
  &      &$\pm$&  0.2 & 0.05 &  1 &  1 & \dots& 0.00 & 0.01 &  0.2 &  10 &   3  \\[1mm]
\hline
\vspace{-4mm}
\end{tabular}
\tablefoot{
\tablefoottext{a}{$T_\mathrm{eff}$ and $\log g$ are
expected to vary over a pulsation cycle in the variable stars (cf.
Table~\ref{observations}), see Sect.~\ref{stellarparameters} for 
further discussion.}
}
\end{table}


\section{Results\label{params}}

\subsection{Stellar parameters\label{stellarparameters}}

Table~\ref{bigtable} summarises the stellar parameters derived from 
the quantitative spectral analysis. This includes the atmospheric parameters 
effective temperature $T_\mathrm{eff}$, surface gravity $\log g$, 
microturbulence $\xi$, projected rotational velocity $v\,\sin\,i$ and
macroturbulent velocity $\zeta$. Additional quantities include the
computed colour excess $E(B-V)$ of the sample stars, their de-reddened 
apparent magnitude
$V_0$, evolutionary masses $M_\mathrm{ev}$, spectroscopic distances
$d_\mathrm{spec}$ and {\sc Hipparcos} distances $d_\mathrm{HIP}$.

\begin{figure}[!t]
\centering
\includegraphics[width=.99\linewidth]{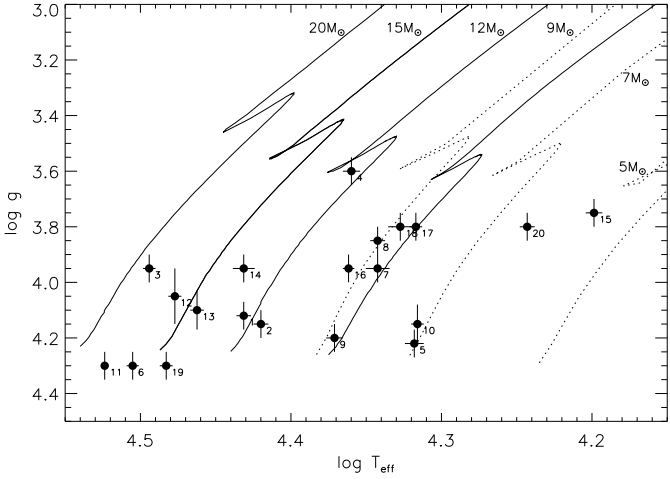}
\caption{The sample stars in the $T_\mathrm{eff}$--$\log g$ plane.
1$\sigma$-uncertainties are shown. Numbering according to
Table~\ref{observations}.
Overlaid are evolution tracks for non-rotating stars of metallicity
$Z$\,=\,0.02 from \citet[full lines]{mm03} and 
\citet[dotted lines]{Schaller92}. See 
Sect.~\ref{stellarparameters} for a discussion.}
\label{tracks}
\end{figure}

Effective temperatures are constrained to 1-2\% and surface gravities
to less than 15\% (1$\sigma$ uncertainties). It is unlikely that these 
values are subject to larger residual systematics\footnote{This is 
strictly valid only for the time of observation, see
Sect.~\ref{obsdata}. The variable stars are expected to show 
(correlated) changes of $T_\mathrm{eff}$ and $\log g$, 
\citep{deRidder02,cale08}, which may exceed the given 
uncertainties in Table~\ref{bigtable}. Average atmospheric parameters 
as derived from the analysis of time-series observations may be more
appropriate in this context, however our approach recovers the time-invariant 
quantities like 
elemental abundances, which is the main topic of the present work.}, 
as they are constrained by the simultaneous match of many {\em independent} 
indicators. Even changes of the underlying physical models, like a use of
hydrostatic non-LTE model atmospheres \citep{n07} or hydrodynamic
non-LTE model atmospheres (NS11), have been shown to have only
small effects (i.e. agreement of stellar
parameters and elemental abundances is obtained with the different
models, within the stated statistical 1$\sigma$-uncertainties and
without systematic trends). 
Overall, this is a major improvement over other literature studies, 
where the uncertainties can amount to $\sim$5-10\% for effective 
temperature 
and $\sim$25\% for surface gravity. In consequence, all quantities 
depending on temperature and gravity (e.g.
reddening, evolutionary masses, spectroscopic distances) also show
reduced uncertainties. A slight degeneracy of line profile            
variations to simultaneous changes of $v\,\sin\,i$ and $\zeta$
prevents one to achieve very tight constraints on these quantities
individually.

A noteworthy result is the finding that {\em one} value of 
microturbulent velocity is derived from the different chemical 
species, after several iterations in all variables of the spectral fitting procedure.
This was often not the case in
previous studies, likely being a consequence of adopting ill-chosen
atmospheric parameters or of shortcomings 
in the modelling, e.g. the assumption of LTE or 
the use of limited sets of atomic data for non-LTE line-formation calculations.

\begin{figure}[!t]
\centering
\includegraphics[width=.98\linewidth]{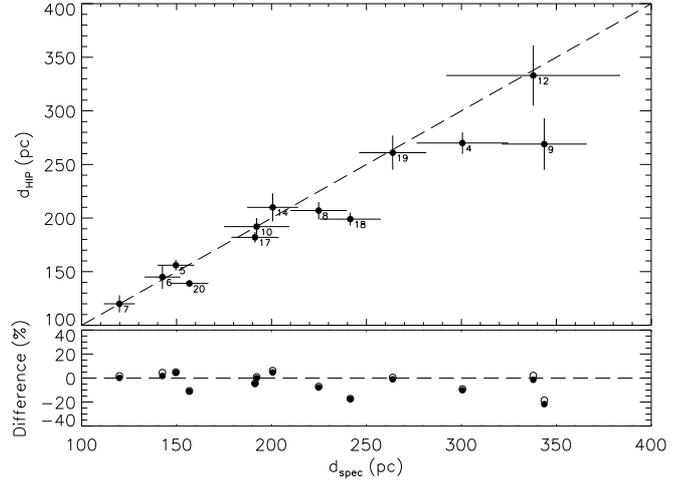}
\caption{Results of the distance determination. Upper panel:
comparison of spectroscopic and {\sc Hipparcos} distances.
1$\sigma$-uncertainties are shown. Numbering according to
Table~\ref{observations}.
The 1:1 relation is indicated by the dashed line.
Lower panel: percent difference of the two distance indicators.
Open circles mark the data if corrected for Lutz-Kelker
bias. See Sect.~\ref{stellarparameters} for a discussion.}
\label{dist}
\end{figure}

A comparison of the resulting model fluxes with the observed spectral
energy distributions (where available) shows good agreement, see
Fig.~4 (available online only). 
Our values for effective temperature derived via multiple independent 
ionization equilibria are thus further verified\footnote{The
opposite approach, using the SEDs as an independent
$T_\mathrm{eff}$-indicator is of limited value because of a much 
lower sensitivity of the method.}. The photometric data was
converted into fluxes using zeropoints of \citet{Bessell98} for Johnson
photometry and of \citet{Heber02} for the 2MASS photometry. All
observed fluxes were de-reddened using an interstellar reddening law
according to \citet{Cardelli89}, adopting colour excesses $E(B-V)$ as
indicated in Fig.~4 and a ratio of total-to-selective
extinction $R_V$\,$=$\,$A_V/E(B-V)=3.1$. Note that the $E(B-V)$ values used in
Fig.~4 may differ slightly (on average by $\sim$0.02\,mag,
which is within the mutual uncertainties) from those in
Table~\ref{bigtable}, which were calculated from the difference between
the observed and the computed {\sc Atlas9} model colour. As IUE
spectrophotometry was not available for all the sample stars, we
decided to use the homogeneously derived $E(B-V)$ data in
Table~\ref{bigtable} to determine~$V_0$.

Evolutionary masses of the sample stars were determined by comparison
of the objects' positions in an effective temperature $T_\mathrm{eff}$ vs. surface gravity $\log g$ diagram with
stellar evolution models from \citet{mm03}, see Fig.~\ref{tracks}. 
Tracks for non-rotating stars with `solar' metallicity $Z$\,=\,0.02 
were adopted. As several stars are less massive than the lower mass
limit of this grid, additional tracks from \citet{Schaller92} were adopted, 
which however show a small offset with respect to the more modern data, see 
the 9\,$M_{\sun}$ models. For consistency, we applied corresponding
offsets to the
lower-mass objects in order to derive homogeneous $M_\mathrm{ev}$
values in Table~\ref{bigtable}. 
Note that some systematic 
offsets will result due to difference between the model ($Z$\,=\,0.02) and 
our derived metallicities ($Z$\,=\,0.014). One consequence will be a 
shift of the zero-age main sequence towards higher gravities, such 
that the high-gravity objects 6, 11 and 19 will fall on the ZAMS.
We neglect this in the following, as the effects on the further
discussion are small.

\begin{table*}[ht!]
\centering
\caption[]{Metal abundances for the sample stars.\\[-6mm]\label{abundances}}
\footnotesize
 \begin{tabular}{@{\hspace{0mm}}rr@{\hspace{0mm}}cccccccccc}
 \noalign{}
\hline
\hline
\#   &HD    & & C & N & O & Ne & Mg & Si & Fe\\ 
\hline\\[-2mm]
 1 &36591 & & 8.33$\pm$0.08\,(30) & 7.75$\pm$0.09\,(61)                  & 8.75$\pm$0.11\,(53) & 8.09$\pm$0.08\,(21)   & 7.58$\pm$0.10\,(6) & 7.50$\pm$0.04\,(5)          & 7.48$\pm$0.11\,(32)\\[.5mm]
 2 &61068 & & 8.27$\pm$0.07\,(23) & 8.00$\pm$0.12\,(61)\tablefootmark{a} & 8.76$\pm$0.09\,(49) & 8.07$\pm$0.11\,(17)   & 7.56$\pm$0.03\,(3) & 7.53$\pm$0.06\,(5)          & 7.51$\pm$0.11\,(28)\\[.5mm]
 3 &63922 & & 8.33$\pm$0.07\,(19) & 7.77$\pm$0.08\,(23)                  & 8.79$\pm$0.10\,(39) & 8.07$\pm$0.07\,(~\,8) & 7.60$\pm$0.01\,(2) & 7.49$\pm$0.12\tablefootmark{b} & 7.51$\pm$0.08\,(~\,9)\\[.5mm]
 4 &74575 & & 8.37$\pm$0.10\,(19) & 7.92$\pm$0.10\,(56)\tablefootmark{a} & 8.79$\pm$0.08\,(45) & 8.05$\pm$0.08\,(12)   & 7.51$\pm$0.10\,(6) & 7.52$\pm$0.12\tablefootmark{b} & 7.51$\pm$0.09\,(27)\\[.5mm]
 5 &122980& & 8.32$\pm$0.09\,(22) & 7.76$\pm$0.08\,(47)                  & 8.72$\pm$0.06\,(52) & 8.07$\pm$0.07\,(14)   & 7.50$\pm$0.05\,(4) & 7.25$\pm$0.04\,(4)\tablefootmark{a}      & 7.44$\pm$0.11\,(27)\\[.5mm]
 6 &149438& & 8.30$\pm$0.12\,(32) & 8.16$\pm$0.12\,(73)\tablefootmark{a} & 8.77$\pm$0.08\,(49) & 8.14$\pm$0.07\,(18)   & 7.62$\pm$0.03\,(3) & 7.52$\pm$0.06\,(2)          & 7.54$\pm$0.09\,(21)\\[.5mm]
 7 &886   & & 8.37$\pm$0.08\,(17) & 7.76$\pm$0.07\,(40)                  & 8.73$\pm$0.11\,(52) & 8.11$\pm$0.08\,(13)   & 7.61$\pm$0.05\,(4) & 7.38$\pm$0.03\,(5)          & 7.51$\pm$0.07\,(30)\\[.5mm]
 8 &29248 & & 8.29$\pm$0.13\,(15) & 7.93$\pm$0.09\,(41)\tablefootmark{a} & 8.78$\pm$0.09\,(47) & 8.07$\pm$0.07\,(12)   & 7.55$\pm$0.08\,(3) & 7.54$\pm$0.06\,(5)          & 7.52$\pm$0.08\,(25)\\[.5mm]
 9 &35299 & & 8.35$\pm$0.09\,(16) & 7.82$\pm$0.08\,(40)                  & 8.84$\pm$0.09\,(52) & 8.07$\pm$0.10\,(14)   & 7.53$\pm$0.06\,(4) & 7.56$\pm$0.05\,(5)          & 7.53$\pm$0.10\,(28)\\[.5mm]  
10 &35708 & & 8.30$\pm$0.09\,(15) & 8.22$\pm$0.07\,(38)\tablefootmark{a} & 8.82$\pm$0.11\,(45) & 8.06$\pm$0.09\,(12)   & 7.65$\pm$0.02\,(4) & 7.51$\pm$0.03\,(5)          & 7.58$\pm$0.06\,(24)\\[.5mm] 
11 &36512 & & 8.35$\pm$0.14\,(19) & 7.79$\pm$0.11\,(22)                  & 8.75$\pm$0.09\,(39) & 8.11$\pm$0.07\,(11)   & 7.50\,(1)          & 7.54$\pm$0.07\,(2)          & 7.53$\pm$0.03\,(~\,3)\\[.5mm]
12 &36822 & & 8.28$\pm$0.14\,(22) & 7.92$\pm$0.10\,(31)\tablefootmark{a} & 8.68$\pm$0.10\,(39) & 8.06$\pm$0.09\,(14)   & 7.54\,(1)          & 7.56$\pm$0.07\,(2)          & 7.52$\pm$0.04\,(~\,9)\\[.5mm] 
13 &36960 & & 8.35$\pm$0.09\,(20) & 7.72$\pm$0.11\,(36)                  & 8.67$\pm$0.08\,(41) & 8.13$\pm$0.11\,(13)   & 7.62\,(1)          & 7.56$\pm$0.07\,(2)          & 7.48$\pm$0.09\,(13)\\[.5mm]
14 &205021& & 8.24$\pm$0.06\,(10) & 8.11$\pm$0.11\,(33)\tablefootmark{a} & 8.64$\pm$0.13\,(44) & 8.17$\pm$0.11\,(10)   & 7.53\,(1)          & 7.50$\pm$0.09\,(2)          & 7.55$\pm$0.10\,(20)\\[.5mm]
15 &209008& & 8.33$\pm$0.09\,(10) & 7.80$\pm$0.11\,(18)                  & 8.80$\pm$0.11\,(21) & 8.02$\pm$0.11\,(14)   & 7.51$\pm$0.07\,(4) & 7.42$\pm$0.04\,(4)          & 7.53$\pm$0.08\,(26)\\[.5mm]
16 &216916& & 8.32$\pm$0.07\,(17) & 7.78$\pm$0.10\,(40)                  & 8.78$\pm$0.08\,(47) & 8.10$\pm$0.11\,(14)   & 7.54$\pm$0.06\,(5) & 7.51$\pm$0.05\,(5)          & 7.50$\pm$0.08\,(21)\\[.5mm]
17 &  3360& & 8.31$\pm$0.08\,(14) & 8.23$\pm$0.07\,(37)\tablefootmark{a} & 8.80$\pm$0.08\,(38) & 8.11$\pm$0.08\,(12)   & 7.56$\pm$0.04\,(3) & 7.60$\pm$0.07\,(5)          & 7.55$\pm$0.07\,(19)\\[.5mm]
18 & 16582& & 8.21$\pm$0.09\,(16) & 8.23$\pm$0.08\,(39)\tablefootmark{a} & 8.79$\pm$0.07\,(44) & 8.05$\pm$0.09\,(12)   & 7.54$\pm$0.05\,(4) & 7.50$\pm$0.05\,(5)          & 7.56$\pm$0.10\,(27)\\[.5mm]
19 & 34816& & 8.38$\pm$0.05\,(10) & 7.81$\pm$0.15\,(29)                  & 8.71$\pm$0.09\,(35) & 8.18$\pm$0.05\,(~\,8) & 7.60\,(1)          & 7.54$\pm$0.06\,(5)          & 7.54$\pm$0.07\,(~\,8)\\[.5mm]
20 &160762& & 8.40$\pm$0.07\,(13) & 7.89$\pm$0.12\,(39)                  & 8.80$\pm$0.09\,(29) & 8.05$\pm$0.07\,(13)   & 7.56$\pm$0.06\,(4) & 7.51$\pm$0.05\,(4)          & 7.51$\pm$0.08\,(22)\\[.5mm]
\hline\\[-5mm]
\end{tabular}
\tablefoot{Uncertainties represent the line-to-line scatter.
The number of lines analysed per element/star is given in brackets.
\tablefoottext{a}{Excluded from the discussion in Table~\ref{abus} 
and Fig.~\ref{hist}. See Sect.~\ref{sect_abundances} for details.}
\tablefoottext{b}{Adopted from PNB08.}}
\end{table*}

Once these parameters are constrained, it is possible to determine
spectroscopic distances $d_\mathrm{spec}$ of the sample stars using a
formula by \citet[][]{Ramspeck01}
\begin{equation}
d_\mathrm{spec}=7.11\times10^4\,\sqrt{M_\mathrm{ev}\,H_{\nu}\,10^{(0.4V_0-\log
g)}}~[{\rm pc}]\,,
\end{equation}
where $M_\mathrm{ev}$ is expressed in units of $M_{\sun}$, the Eddington flux at
the effective wavelength of the $V$ filter $H_\nu$ in
erg\,cm$^{-2}$\,s$^{-1}$\,Hz$^{-1}$ (derived here from the {\sc
Atlas9} models), $V_0$ in mag and $\log g$ in cgs units. The formula uses a
flux calibration of Vega according to \citet{heber84}. The most
crucial input quantity in the distance determination is the surface gravity.
This opens up a possibility to independently verify our $\log g$
determination via comparison of the spectroscopic with {\sc Hipparcos} 
distances as derived from parallaxes $\pi$ from the new reduction of the
{\sc Hipparcos} catalogue \citep{van07}, see Fig.~\ref{dist}.
Lutz-Kelker corrections \citep{LuKe73,Smith03} have not been 
adopted to refine the {\sc Hipparcos} parallaxes, as they should not be 
applied to measurements of {\em individual} stars \citep[p.~87]{van07}. 
Their potential impact is nevertheless rather small for the 
present sample stars, as visualised in Fig.~\ref{dist}.
Overall, good agreement of the spectroscopic and {\sc Hipparcos}
distances is found within the uncertainties, except for 
the objects 9, 18 and 20, which, however, are still within 
the 3$\sigma$ limits. An apparent systematic trend of increasing difference
$d_\mathrm{spec}$\,$-$\,$d_\mathrm{HIP}$ with increasing distance 
(lower panel of Fig.~\ref{dist}) becomes marginal when only one
object, {\#}9, is disregarded, i.e. the regression line is then
compatible with slope and offset zero.

\subsection{Chemical abundances}\label{sect_abundances}
Metal abundances of the sample stars are
summarised in Table~\ref{abundances}, where also the standard
deviation from the line-to-line scatter and the number of analysed
lines are indicated. The values are averages over all lines of a
given species, giving each line in the different ions equal weight.
A precision of the results of better than 25\% is indicated typically. 
The individual line abundances are listed in Table~7
(available online only),
where further details on the line formation calculations are also
given: central wavelengths $\lambda$ of the spectral lines, excitation
energy of the lower level $\chi$, oscillator strengths $\log gf$ and
the accuracy and source of the oscillator strengths.

We deviated from the above procedure for the derivation of the helium 
abundances. Helium is the second most abundant element and therefore
cannot be treated as a trace element: changes in its abundance modify
the mean molecular weight of the atmospheric plasma and thus can affect 
 the atmospheric structure noticeably. 
However, the comparison of the observed with the computed spectra
showed that the sample stars are indeed well-described by a (protosolar)
$\varepsilon$(He)\,=\,10.99$\pm$0.05 (except for object 4, with a --
still compatible -- $\varepsilon$(He)\,=\,10.94$\pm$0.05).

The metal abundances show a small scatter around the average sample 
abundances, which is visualised later, in Fig.~\ref{hist}.
The only peculiarities are enhanced nitrogen abundances in several
sample stars, which can be understood in the framework of mixing of
CN-cycled material into the atmospheric layers \citep[see e.g.][]{p10b}.
The silicon abundance in object 5 is also
conspicuously low, which may be an indicator for the onset of chemical 
differentiation in the otherwise normal star. We decided to
discard these peculiarities from further analysis.

The most relevant sources of systematic uncertainties 
in the chemical abundances that arise from the 
spectral modelling, analysis and observations in our approach 
have already been discussed in a series of papers. We refer to 
the work summarised in Table~\ref{atoms} for estimates of effects and 
systematics due to uncertainties in the atomic input data. Systematics 
due to basic model atmosphere input physics were investigated in
particular by NP07 and NS11, while systematics due to details
of analysis strategies were discussed by \citet{n07} and
\citet{np10,np10b}.

Since our observational material leaves little
room for observational bias we concentrate here only in
quantifying the systematic effects on chemical abundances as 
introduced by uncertainties in the stellar parameters.
We exemplify the effects of independent effective temperature, 
surface gravity and microturbulence variations on oxygen and silicon 
abundances (the most sensitive among the elements in this parameter 
range) for the star $\#$7 (HD\,886, 
$\gamma$\,Peg) in Table~8. Two cases are investigated,
for our and also for typical values in the literature.
Note that this case represents our largest relative
uncertainty in $T_\mathrm{eff}$, one of the largest in
$\Delta\xi$/$\xi$, and typical uncertainty in $\log g$, 
such that the example constrains the {\em maximum} systematic effects 
expected for our sample stars. As correlations exist between
parameters -- e.g. a higher $T_\mathrm{eff}$ implies a higher $\log
g$ in analyses -- the true systematic uncertainties are hard to
quantify in detail. Based on the given example and previous experience
we estimate systematic uncertainties of the elemental abundances in our 
sample stars, accounting for all factors, to be about 0.15\,dex at 
maximum, with the majority  being accurate to within better than 
about 0.10\,dex. In comparison the systematics due to stellar
parameter variations for typical uncertainties from the literature 
are much higher (Table~8), amounting to a factor $\sim$2 
in individual cases. Usually, uncertainties in $T_\mathrm{eff}$ are
the most critical, but note the high sensitivity of the
(rather strong) silicon lines to microturbulence variations.

A test for residual systematics can be made by searching for
trends among the elemental abundances as a function of atmospheric
parameters. Figure~7 (available online only) displays
metal abundances from the present work and from nine additional Orion OB1 
stars of NS11 (analysed in the same manner) 
as a function of $T_\mathrm{eff}$ and $\log g$. All
data points cluster tightly around the average sample values and the
respective 1$\sigma$-error margins (see Table~\ref{abus}), except for
the few cases discussed earlier, mostly for nitrogen. No significant
trends either with $T_\mathrm{eff}$ or $\log g$ are found.

Finally, we want to briefly comment on several potential sources of
systematic uncertainties due to non-standard input physics in the
context of model atmospheres. Potential weak stellar winds present
have a negligible effect on the photospheric line spectrum, as the
effects of the velocity field on the atmospheric stratification become
noticeable only outside the line-formation region. Wind variability by
larger amounts than the mass loss-rates of 
$\dot M$\,$\lesssim$\,10$^{-8}$\,$M_\odot$ typical for B-type
main-sequence stars has no effects on the metal line spectrum
in supergiants \citep[e.g.][]{SchPr08}.
Systematic effects on the abundance analysis due to the presence of 
magnetic fields are also not expected. Only three sample stars have a
confirmed magnetic field, $\beta$\,Cep \citep{henrichs00}, $\zeta$\,Cas
\citep{neiner03} and $\tau$\,Sco \citep{d06}.
To date, none of these shows observational evidence for abundance spots or
vertical chemical stratification.

\subsection{Global spectrum synthesis\label{globalspectrumsynthesis}}
Stellar parameters in Table~\ref{bigtable} in combination with averaged chemical 
abundances in Table~\ref{abundances} were used to compute global synthetic
spectra to visualize the quality 
of the analysis procedure. Overall, excellent agreement is found for all 
stars over the entire observed spectral regions. 
Examples for four stars with spectral types B3\,III 
($\#$15 HD\,209008, 18\,Peg), B2\,IV ($\#$7 HD\,886, $\gamma$\,Peg), 
B1\,V ($\#1$, HD\,36591, HR\,1861) and B0.2\,V ($\#$6 HD\,149438, $\tau$\,Sco) 
are shown in Figs.~8a--11e (available online). 

It is worth to notice that it is relatively simple to achieve reasonable fits 
of models to lower resolution spectra and data at lower S/N over limited 
wavelength regions, i.e. whenever the observational details tend to be washed out.
High-resolution and high-S/N observations spanning a large wavelength range are much
more challenging to be reproduced with synthetic spectra based on
physical models at once. This is feasible only if the models match
the global physics (i.e. the atmospheric structure) and details of 
individual features (i.e. the lines) equally well. We consider our
success as a strong support for the absence of any significant
systematics from our analysis. We thus complement
the probably most comprehensive benchmark test for stellar               
atmosphere modelling of OB stars to date by \citet[at about the upper
$T_\mathrm{eff}$-boundary of the present work]{marcolino09}.

The locations of numerous spectral lines that are considered in our
line-formation computations are indicated in the upper
parts of the individual panels in Figs.~8a--11e 
in order to facilitate an evaluation of their presence/absence 
for a given set of stellar parameters\footnote{More comprehensive line
identifications for early B-type stars are provided e.g. by
\citet{kilian91} for the optical blue spectral region, or by Gummersbach \&
Kaufer, online via {\tt http://www.lsw.uni-heidelberg.de/cgi-bin/websynspec.cgi}}.
These include many more features (also from additional chemical species)
than those analysed quantitatively in
the present work (the latter are marked in the lower parts of the panels). 
We thereby show that even complex blend features like e.g. around 
\ion{He}{i} $\lambda$4120\,{\AA} or around \ion{O}{ii}/\ion{C}{iii}
$\lambda$4650\,{\AA} can be reproduced well. 

There are still some residual minor shortcomings. As can be
seen from the comparison of observation with theory the 
continuum normalisation could be improved globally. However, we have
corrected for this by adjusting the continuum locally 
when analysing individual wavelength regions, such
that this would be rather a cosmetic improvement. Our compromise to fit
the five displayed Balmer lines, H$\epsilon$ to H$\alpha$, results in a
slight mismatch in some of the Balmer line {\em wings} because of the imperfect
normalisation, but the effect is well within the uncertainties of the
$\log g$-determination. The {\em cores} of the Balmer lines for 
the hot star $\tau$\,Sco (and just visible in HR\,1861) are not
matched well because non-LTE effects on the atmospheric structure are
likely to affect the outer photosphere, i.e. the core-formation region
for the strongest lines, gradually with increasing temperature.
Moreover, H$\alpha$ is expected to be influenced by the weak stellar
wind, which is unaccounted for by our hydrostatic modelling. 
The unavailability of realistic broadening data is an issue for some
lines, e.g. for \ion{He}{i}\,$\lambda$3926\,{\AA}.

Several (high-excitation) spectral lines of the elements from
Table~\ref{atoms} and some elements/ions are still not incorporated 
in our non-LTE spectral synthesis. On the other hand, we have included
\ion{Al}{iii} and \ion{S}{ii/iii} in the 
calculations, adopting model atoms of \citet{dufton86} and
\citet{vrancken96}, respectively, and assuming solar abundances 
of \citet{AGSS09}. However, we do not consider these elements 
for the analysis as several shortcomings have been identified in the
model atoms. In consequence, not all of the computed lines of these
two elements give a good 
match to the observed spectra. A thorough verification and improvement
of the model atoms for aluminium and sulphur is beyond the scope of the
present paper, as is the inclusion of the missing lines from other
chemical species. We will report on our efforts for achieving completion of the
non-LTE spectrum synthesis in forthcoming papers.

\setcounter{figure}{11} 
\begin{figure*}[!ht]
\centering
\includegraphics[width=.96\linewidth]{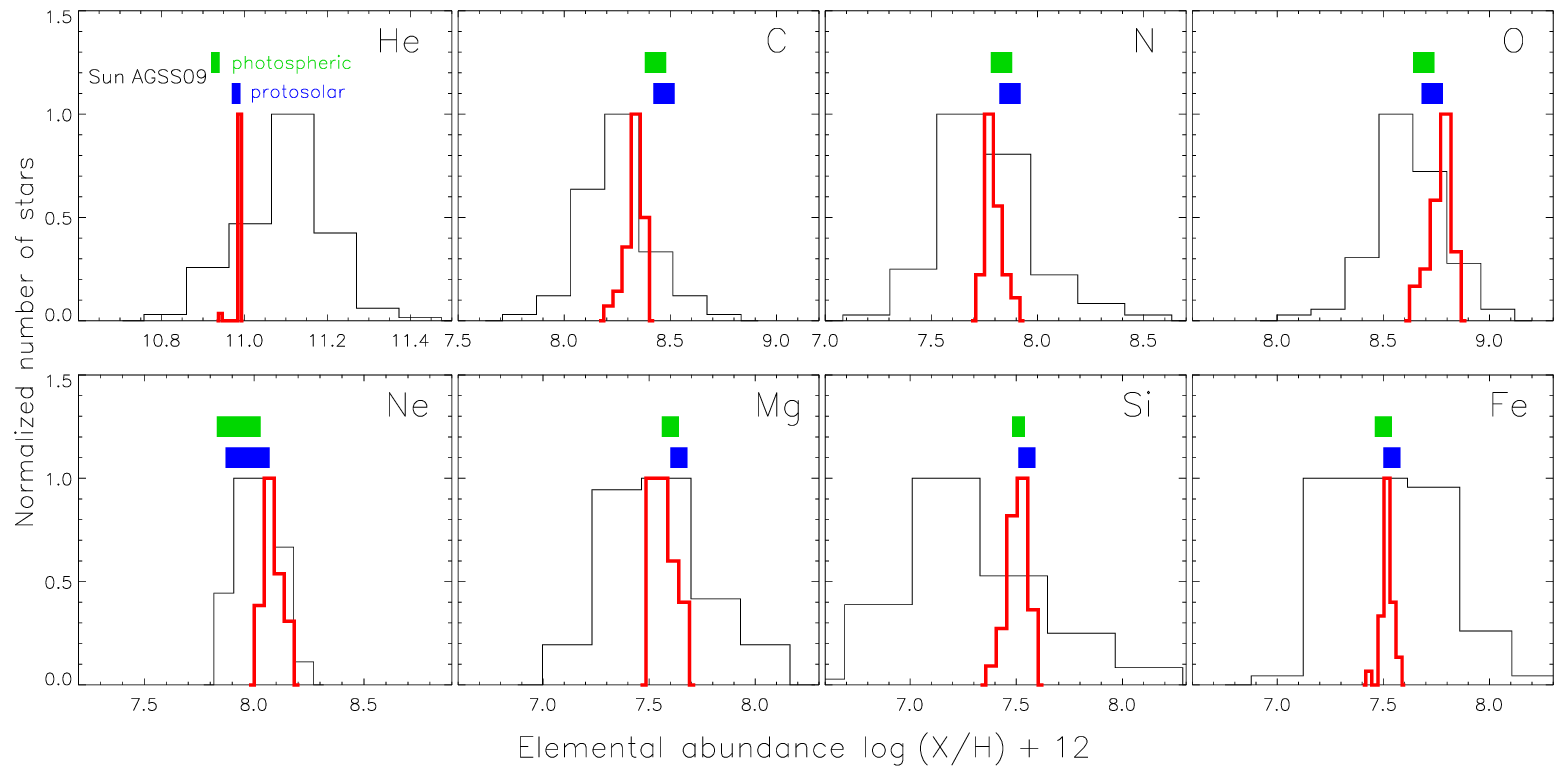}
\caption{Abundance distributions for the astrophysically most relevant
chemical species as derived from early B-type stars in the solar
neighbourhood. Red histograms: present work, establishing the cosmic 
abundance standard. Black histograms: literature data. 
Photospheric and protosolar abundances from AGSS09 are also indicated,
the bars representing the range spanned by the
$\pm$1$\sigma$-uncertainties. See Sect.~\ref{sect_abund} for details.}
\label{hist}
\end{figure*}

\section{Present-day cosmic abundances}\label{sect_abund}
The high degree of chemical homogeneity of the sample stars
(Table~\ref{abundances}) encourages
us to identify the average abundances with the long-sought {\em
cosmic abundance standard} (CAS) for the present-day chemical composition 
of the cosmic matter in the solar neighbourhood. As statistical
significance does matter for such a claim, we seek to compensate for
the seven stars that were removed from our initial sample in order to 
avoid observational biases. We therefore adopt abundance data for 9 
additional early B-type stars from the Ori\,OB1 
association\footnote{Four other stars are in common with the
present work: HD\,35299, HD\,36512, HD\,36591 and HD\,36960. 
The stellar parameters and abundances of NS11 as derived from 
{\sc Fies} spectra agree very well with the present data,
providing another independent consistency check.} (NS11), which 
were derived using the same models and analysis 
techniques as applied here. The stars meet the same observational 
selection criteria ({\sc i-iv} in Sect.~\ref{target_selection}) as our 
core sample.
The spectra were obtained
with {\sc Fies} on the 2.5m Nordic Optical Telescope (La Palma),
covering the wavelength range of 3700--7300\,{\AA} at $R$\,=\,46\,000
and $S/N$\,$>$\,250 \citep{sergio10}. 

Distribution functions for the individual elemental abundances in 
the star sample are displayed in Fig.~\ref{hist} (red histograms). The
data are normalised by the number of sample members, with the maximum
value set to 1. The bin width is chosen as the standard deviation of the
individual distributions. 
Note that only 20 stars are considered in the
histograms for N  -- the atmospheres of nine stars are mixed with
CN-processed material -- and 28 for Si -- one star is Si peculiar -- 
as we are interested in 
the {\em pristine} abundances for constraining the CAS
(see Table~\ref{abundances} for an identification of
the data removed from the discussion here). 

Very tight distributions are found, described by a standard deviation
of typically $\sim$10\%. Mean abundances together with the
standard deviations (of the sample) are given in
Table~\ref{abus}, which summarises also data on present-day abundances
in the solar neighbourhood from other object classes, and the Sun. 
Resulting mass fractions for hydrogen ($X$), helium ($Y$) and 
the metals ($Z$) are indicated in Table~\ref{XYZ}. In addition to the
metals investigated here -- which cover the most abundant ones in the
cosmos --, data for all other metals up to zinc was considered for
constraining $Z$, using solar meteoritic values of \citet{AGSS09},
except for chlorine and argon, where abundances from the Orion nebula were
adopted \citep{esteban04}. Any deviations of this auxiliary data from
the `true' cosmic values will be absorbed by the error margins of $Z$ due to
their small contribution.

This finding of homogeneity is in analogy to PNB08, but on a statistically much more 
significant basis and corrected for slight systematics in the iron
abundances (see Sect.~\ref{models}). The same degree of chemical
homogeneity is recovered as for the {\em gas-phase} of the 
diffuse ISM out to distances of 1.5\,kpc from the Sun \citep{sofia01}, 
see also Table~\ref{abus}, though different absolute
abundance values for many elements are found because of depletion onto dust
grains (see Sect.~\ref{dust}).

\begin{table*}[ht!]
\setlength{\tabcolsep}{.15cm}
\footnotesize
\caption[]{Chemical composition of different object classes in the solar neighbourhood.\\[-6mm]\label{abus}}
\centering
\begin{tabular}{lr@{~~}rcr@{~~}rrrrcr@{~~~}r@{~~~}r}
\hline\\[-2.5mm]
\hline
        & \multicolumn{2}{c}{Cosmic Standard}  & &
	\multicolumn{2}{c}{~~~~~~~~~~~~~~~~~~Orion nebula}              & Young~~~~~                &  \multicolumn{2}{c}{~~~~~~~ISM}&\multicolumn{3}{c}{~~~~~~~~~~~~~~~~~~~~~~~~~~Sun\tablefootmark{k}}\\
\cline{2-3} \cline{5-6} \cline{8-9}  \cline{11-13}
Elem.   & \multicolumn{2}{c}{B stars -- this work\tablefootmark{a}} & & Gas & Dust\tablefootmark{d} & F\&G stars\tablefootmark{e} & Gas & Dust\tablefootmark{j} & & GS98 & AGSS09 & CLSFB10\\
\hline\\[-2mm]
He      & 10.99$\pm$0.01& \ldots      & &10.988$\pm$0.003\tablefootmark{b} & {\ldots} &           \ldots &   \ldots                        &  \ldots     & & \multicolumn{3}{c}{10.93$\pm$0.01}\\[.5mm]
C       &  8.33$\pm$0.04& 214$\pm$20  & &  8.37$\pm$0.03\tablefootmark{c}  & $\sim$0 &    8.55$\pm$0.10 &  7.96$\pm$0.03\tablefootmark{f} &  123$\pm$23 & & 8.52$\pm$0.06 & 8.43$\pm$0.05 & 8.50$\pm$0.06\\[.5mm]
N       &  7.79$\pm$0.04&  62$\pm$6   & &  7.73$\pm$0.09\tablefootmark{b}  & {\ldots} &           \ldots &  7.79$\pm$0.03\tablefootmark{g} & 0$\pm$7  & & 7.92$\pm$0.06 & 7.83$\pm$0.05 & 7.86$\pm$0.12\\[.5mm]
O       &  8.76$\pm$0.05& 575$\pm$66  & &  8.65$\pm$0.03\tablefootmark{c}  & 128$\pm$73 &    8.65$\pm$0.15 &  8.59$\pm$0.01\tablefootmark{h} &  186$\pm$67 & & 8.83$\pm$0.06 & 8.69$\pm$0.05 & 8.76$\pm$0.07\\[.5mm]
Ne      &  8.09$\pm$0.05& 123$\pm$14  & &  8.05$\pm$0.03\tablefootmark{c}  & {\ldots} &           \ldots &   \ldots                        &  \ldots     & & 8.08$\pm$0.06 & 7.93$\pm$0.10 & \ldots \\[.5mm]
Mg      &  7.56$\pm$0.05&36.3$\pm$4.2 & &  6.50:\tablefootmark{c}          & 33.1$\pm$4.2:&    7.63$\pm$0.17 &  6.17$\pm$0.02\tablefootmark{i} &34.8$\pm$4.2 & & 7.58$\pm$0.05 & 7.60$\pm$0.04 & \ldots \\[.5mm]
Si      &  7.50$\pm$0.05&31.6$\pm$3.6 & &  6.50$\pm$0.25\tablefootmark{c}  & 28.4$\pm$4.3 &    7.60$\pm$0.14 &  6.35$\pm$0.05\tablefootmark{i} &29.4$\pm$3.6 & & 7.55$\pm$0.05 & 7.51$\pm$0.03 & \ldots \\[.5mm]
Fe      &  7.52$\pm$0.03&33.1$\pm$2.3 & &  6.0$\pm$0.3\tablefootmark{c}    & 32.1$\pm$2.5 &    7.45$\pm$0.12 &  5.41$\pm$0.04\tablefootmark{i} &32.9$\pm$2.3 & & 7.50$\pm$0.05 & 7.50$\pm$0.04 & 7.52$\pm$0.06\\[.5mm]
\hline
\end{tabular}
\tablefoot{
\tablefoottext{a}{Including nine stars from Orion (NS11), in units of 
$\log ({\rm El}/{\rm H})+12$\,/\,atoms per 10$^6$~H
nuclei -- computed from average star abundances (mean values over all individual
lines {\em per element}, equal weight per line), the uncertainty is
the standard deviation;}
\tablefoottext{b}{\citet{esteban04};}
\tablefoottext{c}{\citet{sergio11};}
\tablefoottext{d}{difference between the cosmic standard and Orion nebula gas-phase abundances, in units of atoms per 10$^6$ H nuclei;}
\tablefoottext{e}{\citet{sofia01};}
\tablefoottext{f}{value determined from strong-line transitions
\citep{sofia11}, which is compatible with data from the analysis of
the [\ion{C}{ii}] 158\,$\mu$m emission \citep{dwek97}. Weak-line studies of
\ion{C}{ii}]\,$\lambda$2325\,{\AA} indicate a higher gas-phase
abundance $\varepsilon$(C)\,=\, 8.11$\pm$0.07 \citep{sofia04}, which
corresponds to 84$\pm$28\,ppm of carbon locked up in dust;}
\tablefoottext{g}{\citet{meyer97}, corrected accordingly to \citet{jensen07};}
\tablefoottext{h}{\citet{cartledge04};}
\tablefoottext{i}{\citet{cartledge06}. The uncertainty in the ISM
gas-phase abundances is the standard error of the mean;}
\tablefoottext{j}{difference between the cosmic standard and ISM gas-phase abundances, in units of atoms per 10$^6$ H nuclei;}
\tablefoottext{k}{photospheric values of \citet[GS98]{gs98},
\citet[AGSS09]{AGSS09} and \citet[CLSFG10]{Caffau10}.}} 
\end{table*}

However, the finding is at odds with practically all previous work on
early B-type stars in the solar neighbourhood. We concentrate on literature 
data from homogeneously analysed samples of more than 10 stars 
for the comparison with the present work, as a comprehensive review
of all available data is beyond the scope of this paper. Therefore, 
abundances from the studies of
\citet{kilian92,kilian94}, 
\citet{gies92} -- excluding bright giants and supergiants --, 
\citet{cunha94}, 
\citet{daflon99,daflon01a,daflon01b, daflon03}, 
\citet{chl06}, \citet{mb08} and 
\citet{lyubimkov04,lyubimkov05} 
are adopted, essentially applying the same distance cut as in our sample
selection. These abundances were derived from high-resolution spectra 
using comparable non-LTE techniques as utilised here: either hybrid
non-LTE or full non-LTE modelling under consideration of metal
line blanketing. The only exception are iron abundances, which were 
determined in LTE in these studies.
The combined distribution functions\footnote{Note that many stars
were subject to two or more of these independent investigations. No attempt
is made to single out these cases: each analysis is considered with
equal weight in the histograms.} for the individual elemental abundances
from the literature are also displayed in Fig.~\ref{hist} (black
histograms). Much broader distributions are
indicated, with typical standard deviations of about 0.2\,dex. This is a
factor $\sim$5 larger than in the present work, {\em despite our sample
is a representative sub-set of the previously investigated stars}, see
Sect.~\ref{target_selection}.

\begin{table}[t!]
\footnotesize
\caption[]{Mass fractions for hydrogen, helium and metals.\\[-6mm]\label{XYZ}}
\centering
\begin{tabular}{lccrrr}
\hline\\[-2.5mm]
\hline
  & Cosmic Standard & & \multicolumn{3}{c}{Sun -- photospheric values}\\
\cline{2-2} \cline{4-6}  
  & B stars -- this work & & GS98  & AGSS09 & CLSFB10\\[.5mm]
  \hline
$X$ & 0.710           & & 0.735 & 0.7381 & 0.7321\\
$Y$ & 0.276           & & 0.248 & 0.2485 & 0.2526\\
$Z$ & 0.014$\pm$0.002 & & 0.017 & 0.0134 & 0.0153\\
\hline
\end{tabular}
\end{table}

It is extremely difficult to trace the discrepancies in stellar parameters 
and elemental abundances of individual stars from the various 
investigations, as differences exist at all levels: 
the quality of the observational material, the methodologies of stellar 
parameter determination, the choice of analysed lines, the input atomic 
data, the computer codes and assumptions used for the modelling, 
among many other details 
that differ from study to study. A combination of several factors
is most likely responsible for the discrepancies. We do 
not aim at resolving these
discrepancies in detail case by case as little can be learned in terms
of the objective of the present work. Moreover, the overall picture can 
actually be understood rather well from some basic considerations.

Any abundance determination using a method with finite precision will
yield an abundance distribution with a larger dispersion than the
true one. Broad distributions like those derived from the 
literature data {\em can} result from underlying tight distributions.
Actually, the finding of such broad distributions is {\em expected},
given that the statistical 1$\sigma$-uncertainties in the literature data alone
can amount to 0.2 to 0.3\,dex and systematic uncertainties
result in shifts of the derived abundance distributions of the individual
studies relative to each other (see Fig.~2 of PNB08 for a visualisation). 
A full spread of the literature data over 1\,dex and a
shift in mean abundances is therefore hardly surprising.

Finally, Fig.~\ref{hist} allows also a comparison of the abundance
distributions for the B-stars with the solar standard to be made.
Photospheric abundances of \citet{AGSS09} are chosen as a
representative for the type of data that is typically adopted in all
kinds of astrophysical literature, and protosolar values from the same
source as a representative for the bulk composition of the Sun
(correcting for a $\sim$0.04\,dex depletion of the photosphere due to
diffusion). Interestingly, similarities are found for some elements and 
differences for others, which will be discussed further in
Sect.~\ref{sun-foreigner}. Note that the 1$\sigma$-uncertainties of the 
solar abundances are about the same as the B-star sample standard deviations. 
{\em Technically, the fact that a larger number of B-stars is considered
for the determination of the CAS -- in contrast to one star defining
the solar standard --, means that the uncertainties of the
CAS-values can be expressed via the respective standard error of the mean, 
which amounts to 0.01\,dex for all metals studied here}. However, we
prefer to assign the standard deviation of the sample as {\em conservative}
error margins for the CAS. 

\begin{figure}[t!]
\centering
\includegraphics[width=.99\linewidth]{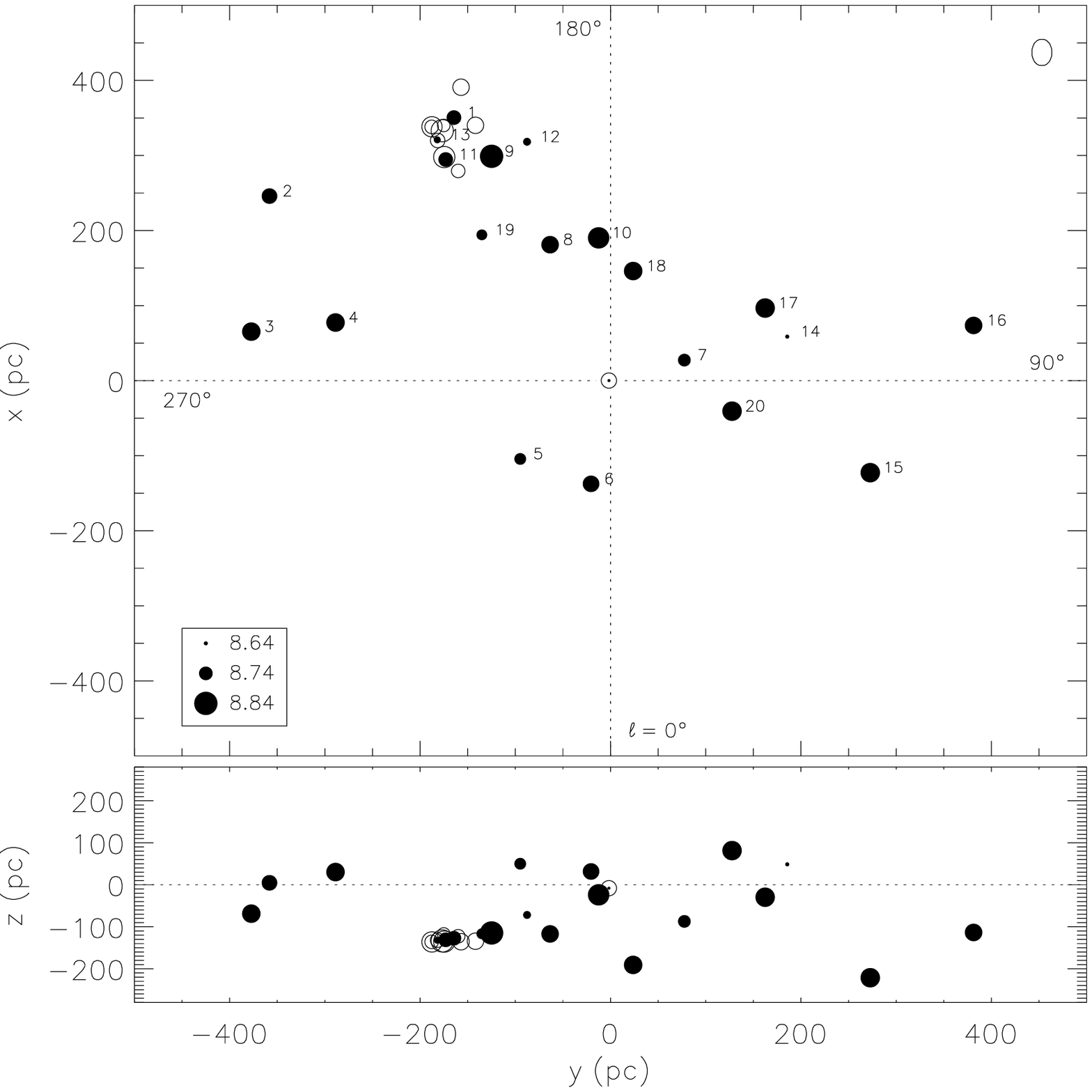} 
\caption{
Spatial distribution of oxygen abundances in the sample stars, in
analogy to Fig.~\ref{distribution}. Open circles represent objects
from NS11. The symbol size encodes the abundance according to
the figure legend.
}
\label{abu_dist}
\end{figure}

In our quest to reduce systematic errors to a minimum we cannot ignore
possible bias introduced by other factors than stellar atmospheres
alone. When considering an extended region like the solar
neighbourhood (in our definition), effects from Galactic chemical evolution 
 may also come into play. The presence of Galactic abundance gradients
implies a decrease of metal abundances with increasing Galactocentric
radius. To verify this, we checked for correlations of
the stellar abundances with the spatial positions of the stars. An
example for oxygen is shown in Fig.~\ref{abu_dist}. No correlations
are found, neither with Galactocentric distance nor with distance from
the Galactic plane. We conclude that signatures of Galactic chemical evolution
are insignificant on scales of $\sim$500\,pc, providing no bias to the CAS
on the level of precision achieved with our analysis methodology. 

In summary, it emerges from our previous discussion that the drastic 
reduction of many systematic uncertainties in our analysis procedure
is the key for understanding the derived small dispersion 
in the elemental abundances of the sample stars. For the first time the 
true abundance distributions of the young stellar population in our
Galactic vicinity are approximated, which appear intrinsically tight. 
At the same time, the overall 
match of a large number of independent observational
constraints and the successful passing of numerous tests for remaining
biases puts confidence in the accuracy of the results. This allows an 
accurate and precise cosmic abundance standard to be established.
We expect that the true abundance distributions for our B-star 
sample will be in fact tighter than derived here because they are subject
to analysis with a methodology of finite precision. Therefore,
{\em our study gives only a upper limit on the true degree of 
chemical homogeneity of the present-day cosmic matter in the solar 
neighbourhood}. Given the current state of input physics for the models,
it will be highly costly to improve the analysis inventory to a degree
where much tighter constraints can be achieved.

\section{Implications}\label{sect_impl} 
In the following we want to investigate what implications the use of
the cosmic abundance standard instead of the solar standard has on 
various astrophysical fields. We concentrate on the impact of our
sample data and the resulting CAS values for the evolution of
massive stars, for ISM science and for Galactic chemical evolution.
Finally, we briefly comment on the origin of the Sun and its relation to 
its present Galactic neighbourhood.

\begin{figure}[!t]
\centering
\includegraphics[width=.99\linewidth]{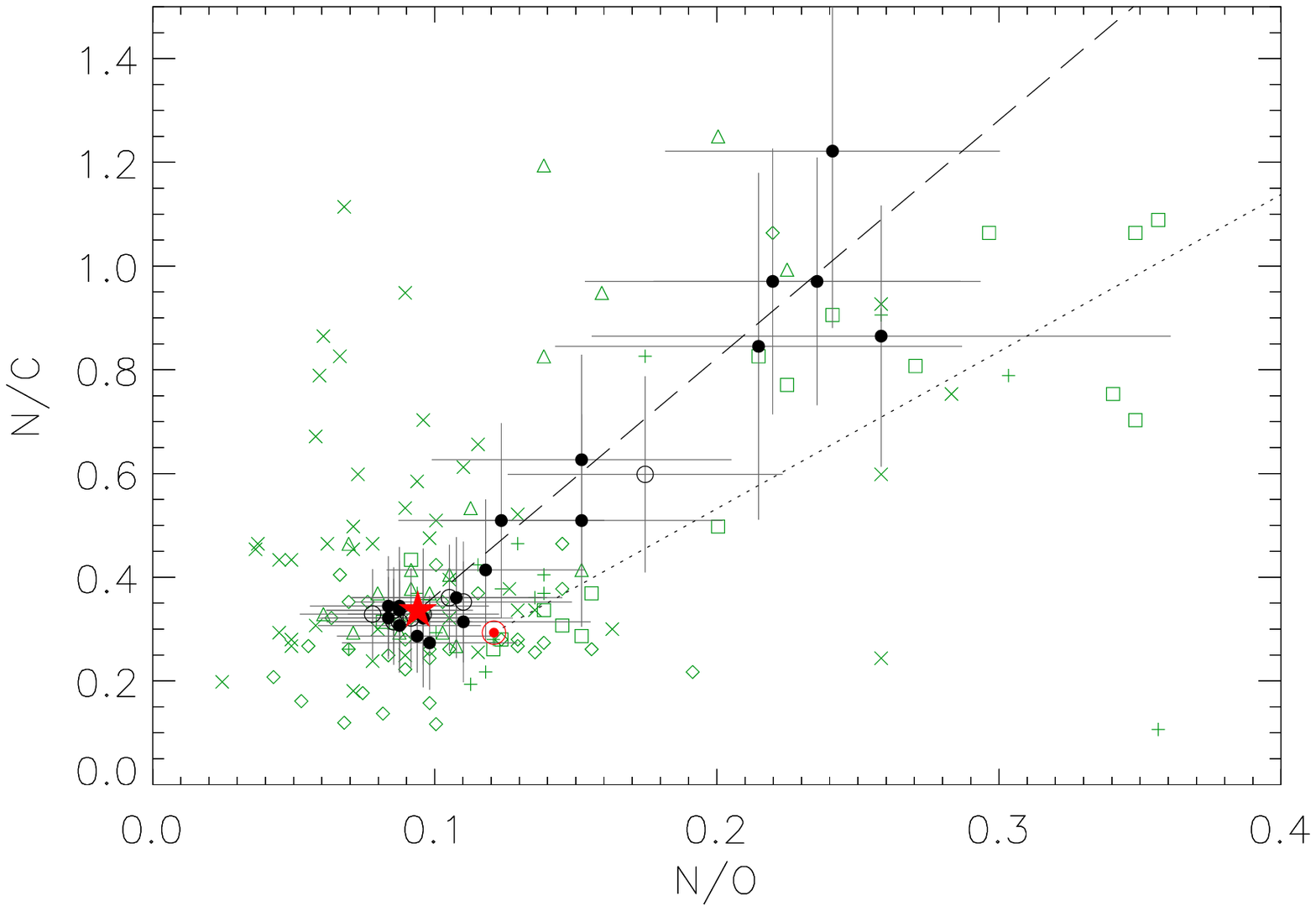}
\caption{
Observational constraints on 
mixing of CNO-burning products in massive stars. Mass ratios N/C over N/O 
are displayed. Black dots: present data; black circles: 9 additional 
objects from NS11. Data from previous non-LTE analyses are
given as grey symbols (green online) --  
plus signs: \citet{kilian92}; triangles: \citet{gies92};
diamonds: \citet{cunha94}, \citet{daflon99,daflon01a,daflon01b};
squares: \citet{morel08}; crosses: \citet{hunter09}.
The predicted nuclear paths assuming initial CAS abundances (red star)
and solar abundances ($\odot$) of \cite{AGSS09} are
indicated by the long-dashed and dotted lines, respectively.
Statistical uncertainties of 0.2\,dex in each element -- 
which are typical for previous literature data -- result in an error 
bars more than twice as large than those from the present data.}
\label{cno_mix}
\end{figure}

\subsection{Stellar evolution: the initial chemical composition}\label{stellarevolution}
The initial chemical composition has a profound influence on the
structure and evolution of stars because of its effect on opacities
and mean molecular weight. We have addressed the effect of a
metallicity reduced from the so far canonical value $Z_\odot$\,=\,0.02
to $Z_{\rm CAS}$\,=\,0.014 in the discussion of Fig.~\ref{tracks}. 
The shift of the ZAMS towards higher gravities will also be
accompanied by a shift of the evolution tracks towards
higher effective temperatures. But, would there be a significant effect
if the CAS or modern solar abundances at about the same $Z$ (see
Table~\ref{XYZ}) were used? In terms of the position of the tracks in
the Hertzsprung-Russell diagram -- probably not; however, in terms of
observable tracers for rotational mixing, certainly yes.

Energy production in massive stars is governed by the CNO cycles
throughout most of their lifetime and the nuclear-processed material 
may reach their surface layers through rotational mixing
already during their main sequence phase \citep[e.g.][]{mame00,hela00},
opening up a very powerful diagnostic to test models of stellar evolution. 
The changes of the CNO surface abundances reflect the actions of the
dominating CN-cycle initially, following a well-defined nuclear path.
This can be analytically approximated by a straight line in the 
diagnostic N/C--N/O-diagram (see Fig.~\ref{cno_mix}), 
with a slope defined solely by the initial CNO abundances \citep{p10b}, 
regardless of the mass, initial velocity or other model details.

The present analysis of our sample stars (and the additional nine
objects from NS11) facilitates the predicted trend 
to be recovered for a statistically significant sample for the first
time, see Fig.~\ref{cno_mix}. Most of our objects
cluster around the pristine CAS values, i.e. they are unmixed, 
while about 1/3 of the stars show a mixing signature of varying
magnitude, following the predicted nuclear path with d(N/C)/d(N/O)\,=\,4.6 
(for initial CAS abundances) tightly. Stellar evolution models based e.g.
on the solar values by AGSS09 would predict a different nuclear path 
(with slope $\sim$3.0) despite a rather similar 
bulk metallicity. 

The large scatter found by previous non-LTE analyses of early B-type 
stars in the solar neighbourhood (many objects 
are in common with our sample) and additionally in three Galactic clusters
is also likely a consequence of
the lower accuracy and precision achieved in these studies, as argued
in Sect.~\ref{sect_abund}. Most data points are in 
fact consistent with the predictions because of the larger error bars,
but they are of limited use for testing stellar evolution 
models {\em stringently}.

Further consequences of the use of different individual abundances will 
be modified yields. As these are key input for 
Galactic chemical evolution models, they have to be determined as realistically as possible.

\begin{figure}[!t]
\centering
\includegraphics[width=.95\linewidth]{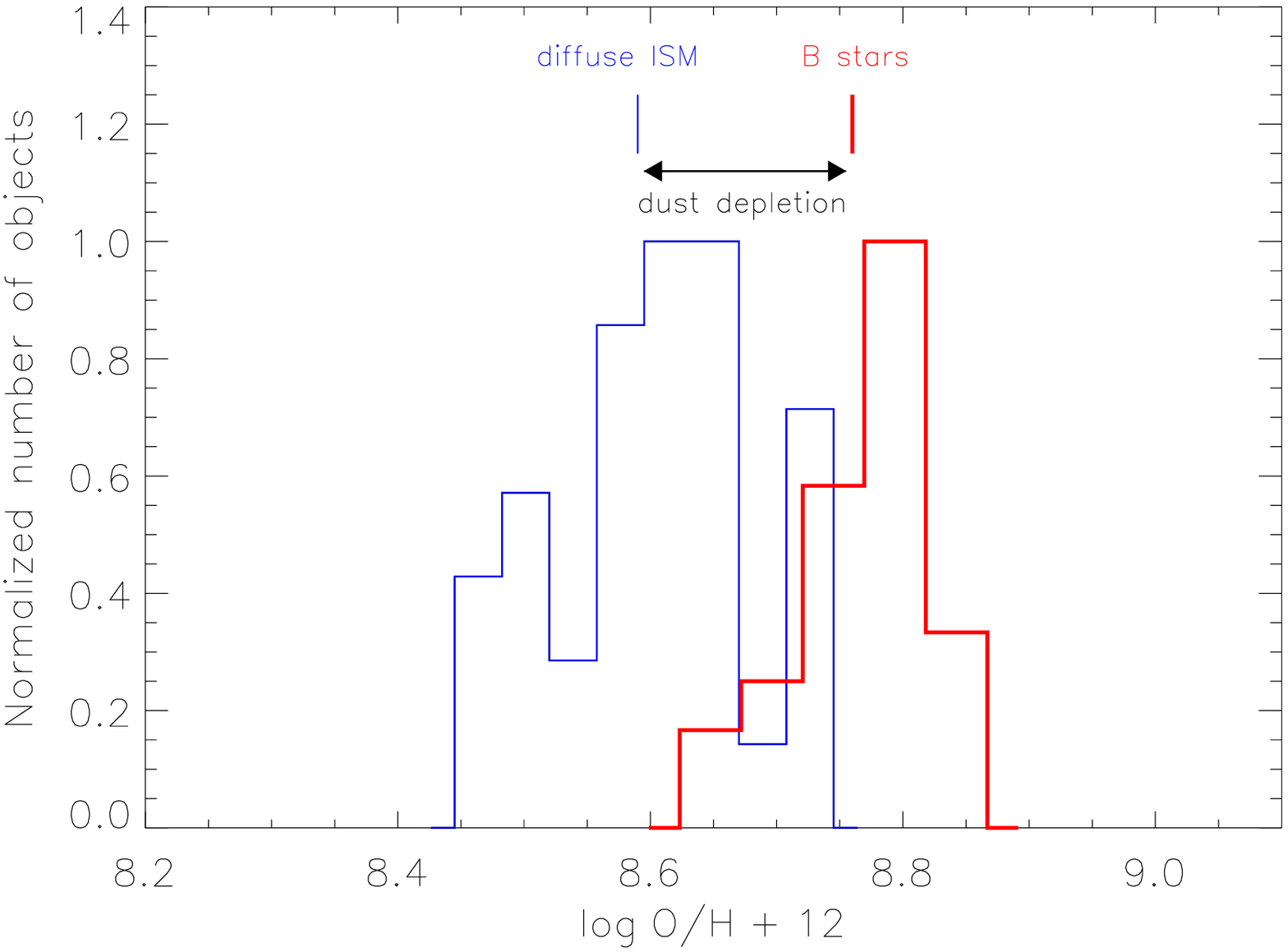}
\caption{
Comparison of the stellar O abundance distribution as derived in the
present work (red histogram) with gas-phase abundances along different
sightlines of the diffuse
ISM \citep[thin blue histogram,][]{cartledge04}, in analogy to
Fig.~\ref{hist}. Mean abundances are indicated. 
}
\label{dust_plot}
\end{figure}

\subsection{Chemical homogeneity: early B-type stars vs. ISM\label{ISMhomogeneity}}
From studies of interstellar absorption lines of the cold gas it is known for a
long time that the local ISM out to 1.5\,kpc from the Sun is
chemically homogeneous, to the 10\%-level \citep{sofia01}. 
This can be understood as a natural consequence of turbulent mixing 
acting on all scales, which is due to the large density 
variations of the gas, generated by a complex interaction of many
factors like momentum injection by stellar winds and supernova shocks, 
magnetic fields and self-gravity.
Theoretical investigations of metal abundance fluctuations in the ISM 
have until recently been based on order of magnitude arguments involving 
characteristic spatial scales and timescales for various turbulent mixing 
processes, see e.g. \citet{rk95}, and \citet{se04} for a review.
More recently, sophisticated 3D (magneto-)hydrodynamic simulations of
the local ISM at high resolution \citep[e.g.][]{ab07} support the view
of an efficient mixing of metals at wide ranges of scales, driven          
by turbulence.  

The young stars in the solar neighbourhood are expected to follow the
chemical characteristics of the matter from which they were formed.
Indeed, the present study shows that, independent of the location 
of the sample stars in the solar neighbourhood (see Fig.~\ref{abu_dist}) 
-- whether they reside in OB associations or 
in the field --, and also independent of their mass ($\sim$6 to 20 M$_{\odot}$) 
and hence their life-time ($\sim$5$\times$10$^7$ to 5$\times$10$^6$\,yr, 
respectively), all stars show practically the same chemical composition.
The fluctuations around the mean are $\sim$10\%, and probably less. 
This {\em independent} verification of the results from investigations of the
ISM gas is achieved for the first time. 

The huge advantage of studying
early-type stars is that the {\em entire} metal content can be
determined using quantitative spectroscopy, with no material hidden in
an observationally inaccessible reservoir like the dust-phase in the ISM.
This opens up the possibility to determine the chemical composition of
the dust in an indirect way, see the next sub-section. 

Moreover, our results put constraints on the injection and mixing                   
timescales of metals in the local Galactic ISM, and therefore 
on the hydrodynamics of the ISM. It appears that fresh 
nucleosynthesis products from supernovae and AGB (super-)winds 
or infall of pristine material onto the Galactic disk are unlikely to lead to 
a noticeable (at the level of the present abundance determination 
precision) local enrichment or depletion in 
a high-metallicity environment like the solar neighbourhood
over a timescale of several 10$^7$\,yr. Or, in other words,
the interaction of hydrodynamic mixing on the one hand and viscosity and 
molecular diffusion on the other is highly efficient, such that
the medium is homogenised~quickly.

\subsection{Dust composition of the local ISM}\label{dust}
An important open question in our understanding of the ISM is the
chemical composition of the dust particles. The amount of metals
incorporated into dust cannot be derived directly from
observations. Only an indirect determination is feasible, 
from the comparison of the ISM gas-phase abundances and a suitable
reference that is unaffected by depletion onto dust grains. There is
an ongoing debate which kind of objects provide the reference suited
best for the comparison: young B-type stars, young F\&G-type stars, or
the Sun \citep{sofia01}. 

\begin{figure*}[!t]
\includegraphics[width=.495\linewidth]{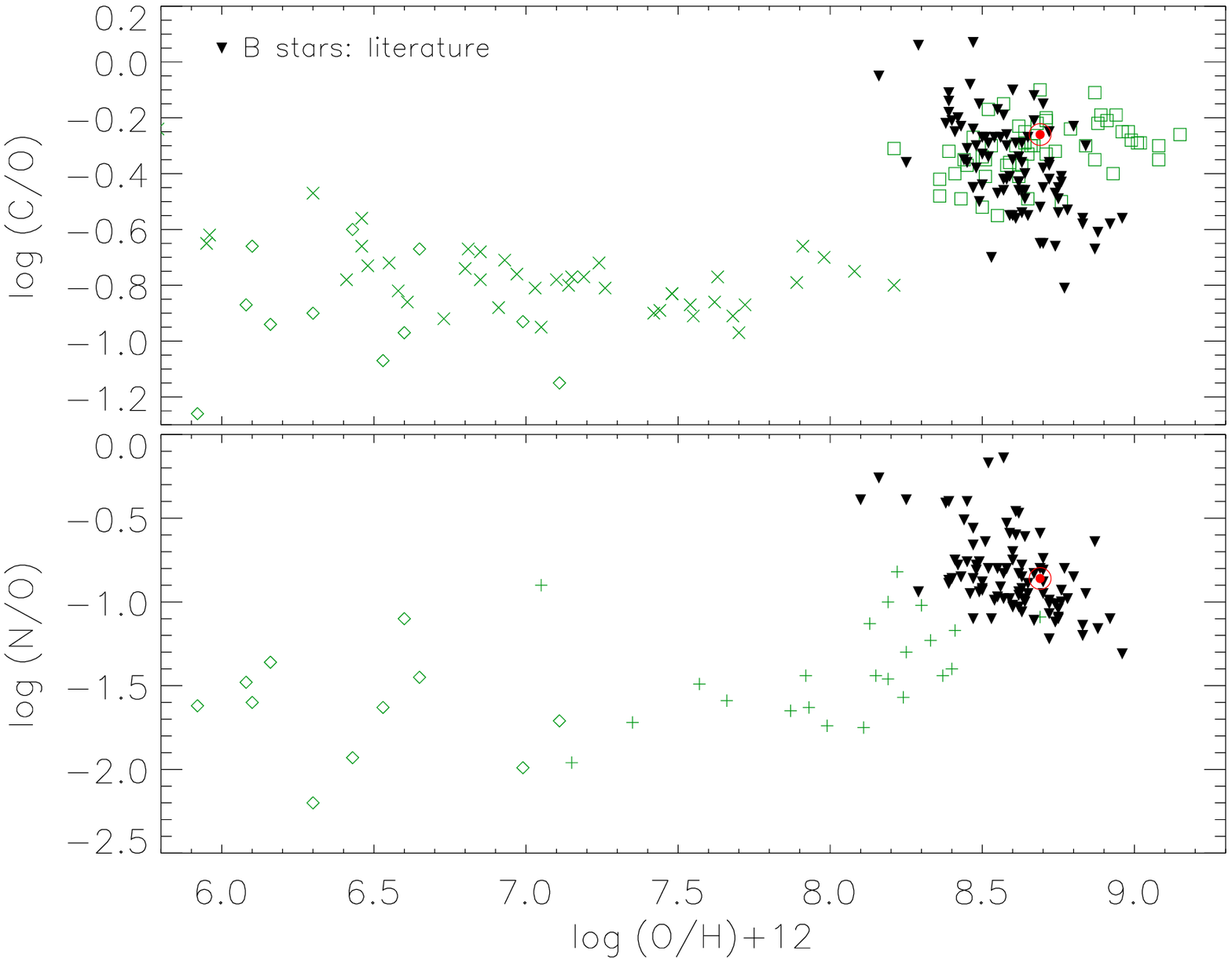}
\hfill
\includegraphics[width=.495\linewidth]{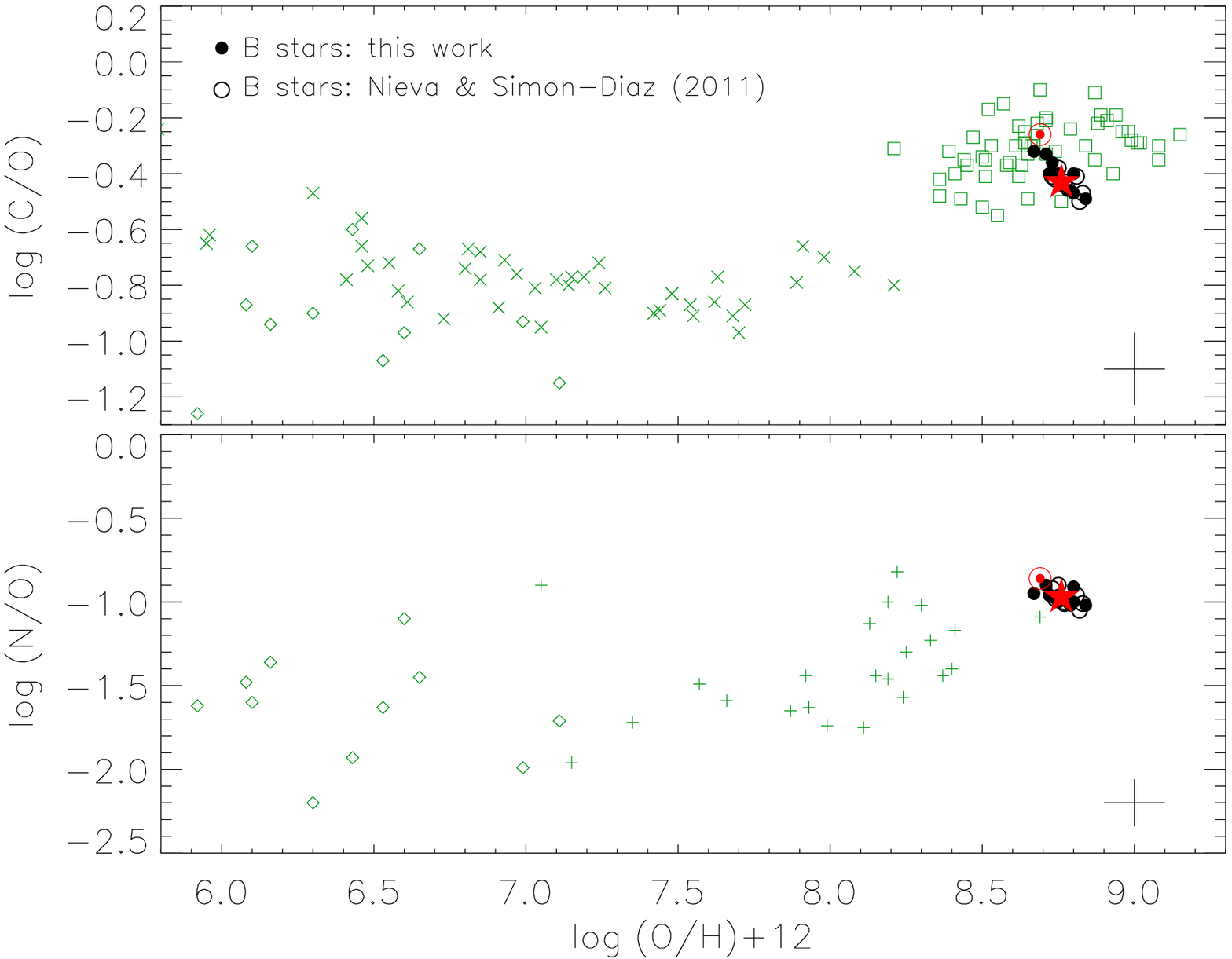}
\caption{Observational constraints on the chemical evolution of Galactic
CNO abundances: abundance ratios log\,(C/O) and log\,(N/O) vs. 
O abundance. 
{\em Left panels}: black triangles: early B-stars from the literature
extracted from Fig.~\ref{hist}. 
Data for low-mass stars are displayed as grey symbols 
(green in the online edition) -- squares: \citet[solar-type
dwarfs]{gustafsson99}; crosses: \citet[solar-type dwarfs and
subgiants]{fabbian09}; diamonds: \citet[unmixed cool giants]{spite05};
plus signs: \citet[unevolved solar-type stars]{israelian04}.
Solar abundance ratios of \cite{AGSS09} are also indicated ($\odot$).
{\em Right panels}: black dots: unmixed objects from the present work, 
black circles: unmixed stars from NS11. Data from the literature
like in the left panels. The CAS is also indicated (red star).
Error bars (statistical 1$\sigma$-uncertainties) typical for 
individual stars in the present study are shown.} 
\label{cno_evolution}
\end{figure*}

We argue here that the CAS values as established from the analysis of
early B-type stars provide the long-sought reference, unprecedented in
precision and accuracy. Most notable is that the same degree of chemical
homogeneity is found for both, the CAS reference and the gas-phase
abundances of the diffuse ISM (see Table~\ref{abus}). The abundance
distributions are very similar, see Fig.~\ref{dust_plot}
for the example of oxygen \citep[the gas-phase abundance distribution is
based on 37 diffuse sightlines of][]{cartledge04}.
From this it
follows immediately that the dust-phase is chemically also rather homogeneous,
what can be expected if mixing processes are highly efficient within
the ISM (Sect.~\ref{ISMhomogeneity}). 

We derive a similar chemical composition for the dust grains as
PNB08, but at much higher statistical significance and with
two exceptions (see Table~\ref{abus}). The Fe abundance is higher due to 
the identification and 
removal of residual systematics in the line-formation computations for that
element (Sect.~\ref{models}). For carbon, a recent investigation by
\citet{sofia11} raises doubts about the precision of weak-line
analyses based on the \ion{C}{ii}]\,$\lambda$2325\,{\AA} transition
(see \citet{sofia04} for a discussion),
possibly related to a systematically underestimated oscillator strength.
We therefore adopt a mean abundance from five sightlines of the strong-line
analysis of \citet{sofia11}, which is compatible with data from the
[\ion{C}{ii}]\,$\lambda$158\,$\mu$m line by \citet{dwek97}.

Overall, the results indicate a
silicate/oxide-rich and relatively carbon-poor composition for the local
ISM gas phase. In particular, the CAS provides sufficient oxygen to
sustain the values required by magnesium, silicon and iron to be locked up by vast
majority in silicates (plus a small fraction in metal oxides) in the diffuse ISM. 
Using a reasonable dust composition \citep{draine03} this amounts to 
about 140--150\,ppm of oxygen for the given abundances of the
refractory elements in the dust, in good agreement with the
observed value (Table~\ref{abus}), with some additional oxygen
possibly incorporated in organic compound material. 
About 60\% of the total carbon resides in
the dust phase. Despite a higher abundance of carbon is found in the dust
in absolute terms relative to PNB08, this falls still somewhat short of
the demands of most dust models, see e.g. the discussions by 
\citet{snow95} and \citet{zubko04}. 

Finally, we want to comment on the cosmic
abundance standard in the context of gas- and dust-phase abundances in
the Orion nebula. Further information for the general picture can
be gained under the -- not unlikely -- assumption that the Orion giant 
molecular cloud formed out of material typical for the diffuse ISM,
with subsequent chemical processing taking place in the cloud core.

A comparison of the CAS with gas-phase abundances 
\citep{esteban04,sergio11} implies that the {\em \ion{H}{ii} region is
devoid of carbonaceous dust}. From observations of the
photodissociation region in the Orion nebula it is known that 
polycyclic aromatic hydrocarbons (PAHs) disappear as the gas becomes ionized
\citep[e.g.][]{tielens08}.
Our results imply that photoevaporation affects all carbon-bearing dust 
particles in a similar manner, indicating that {\em little carbon was
incorporated initially in graphite} (the most stable form of
carbon under interstellar conditions), in line with the findings of \citet{amari90}
from laboratory studies of meteorites. Also, amorphous carbon dust 
grains\footnote{Amorphous carbon is considered the predominant grain 
material produced by C-type asymptotic giant branch stars, the main source of carbonaceous dust
\citep[e.g.][]{wallerstein98}.} are either efficiently destroyed inside 
the ionized region, or
they were a minority species initially as well. On the other hand, there is only 
weak evidence for the destruction of silicate grains 
from the numbers in Table~\ref{abus} -- the abundances of oxygen 
and of the refractory elements in the Orion nebula dust-phase are 
compatible with the ISM dust data within the (large) error bars. 
Unfortunately, a more direct verification of the carbon-poorness of
dust within the \ion{H}{ii} region via e.g. the extinction properties is 
complicated, as most of the extinction towards Orion occurs in the neutral medium
surrounding the \ion{H}{ii} gas \citep{baldwin91}. 

The combined evidence from abundances in the ISM and in the Orion 
\ion{H}{ii} region indicates that {\em dust models considering silicates,
PAHs, organic refractory material and possibly amorphous carbon, but
not graphite, should be investigated more closely}. Models in analogy
to the COMP-NC-type or COMP-AC-type models of \citet{zubko04} look 
highly promising for future studies in view of abundance demands and the 
ability to match other observational constraints like extinction
and emission properties of the dust. We are confident that the tight
observational constraints provided by the CAS will facilitate a better
understanding of the nature of dust and grain structure to be developed.

\subsection{Galactochemical evolution: present-day abundances\label{GCE}}
Nucleosynthesis in successive generations of stars has enriched
the cosmic matter with heavy elements ever since the first Population III
stars were born. Studies of various objects like (Galactic and
extragalactic) stars and \ion{H}{ii} regions, or the ISM in damped
Ly$\alpha$ (DLAs) systems allow the cosmic
enrichment history to be traced and the specific production sites 
of individual elements to be constrained.   
The CAS provides valuable input for the comparison of models with
observations, as it marks the {\em present-day endpoint of galactochemical
evolution}, in particular for a typical spiral galaxy like the Milky Way.
We put CAS values into the context of the evolution of the five most important
chemical species, the light elements CNO, magnesium as a typical
tracer of the $\alpha$-process in core-collapse supernovae and iron as
tracer of iron-peak nucleosynthesis in supernovae of type Ia.

\begin{figure*}[!t]
\includegraphics[width=.495\linewidth]{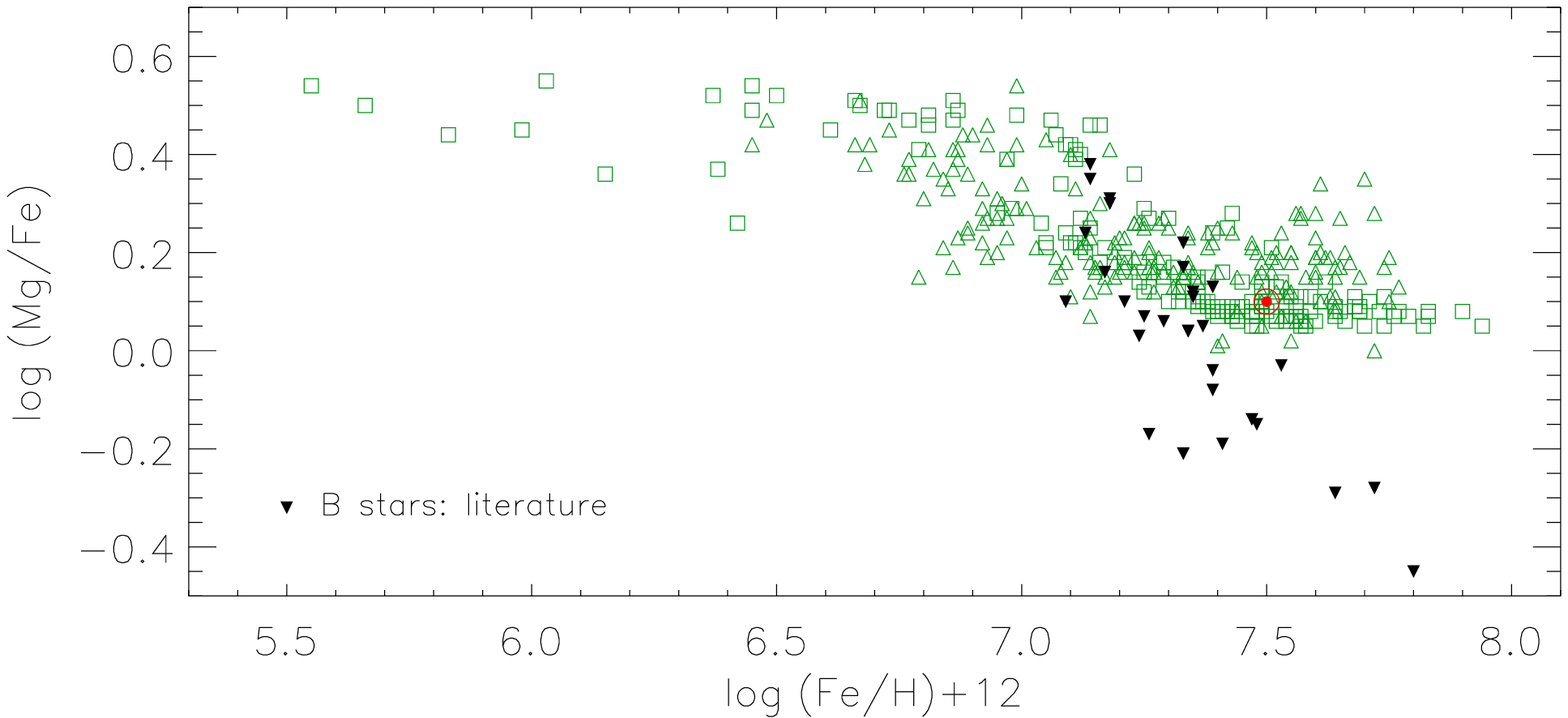}
\hfill
\includegraphics[width=.495\linewidth]{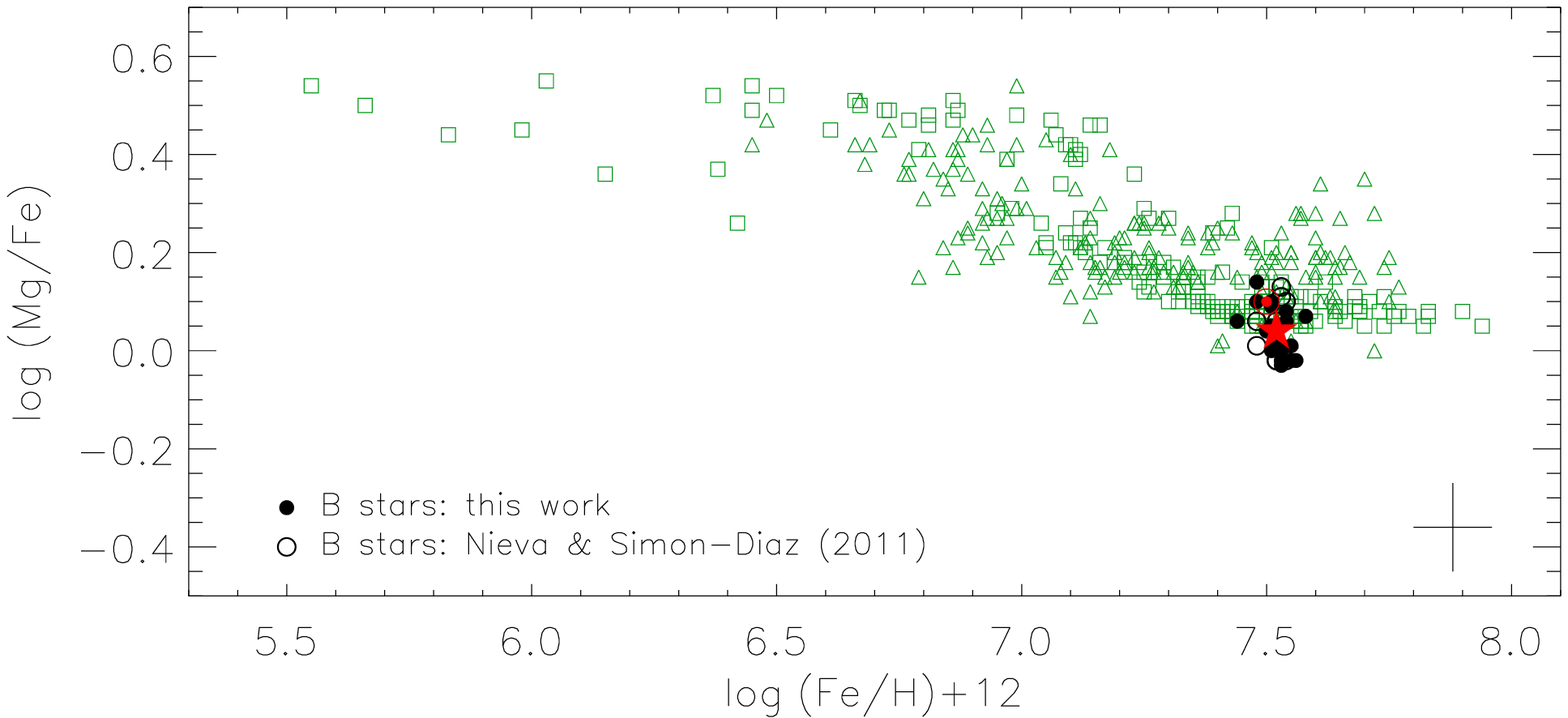}
\caption{
Evolution of 
Galactic $\alpha$-process vs. iron-group elements: log\,(Mg/Fe)  
as a function of Fe abundance. 
Literature data on solar-type stars are displayed as grey symbols
(green online) -- 
squares: \citet{fuhrmann98,fuhrmann04}; 
triangles: \citet{EAGLNT93}. All other symbols as in
Fig.~\ref{cno_evolution}.}
\label{mgfe_evolution}
\end{figure*}

Observational constraints on the interlinked evolution of CNO are
displayed in Fig.~\ref{cno_evolution}. Abundances from early B-type
stars are compared to data from solar-type stars in the Galactic thin and thick
disk, and in the halo. 
Typical statistical accuracies for abundances in solar-type stars are 
$\sim$0.05-0.10\,dex for LTE analyses, similar to the present work. 
Note, however, that systematic uncertainties 
due to non-LTE and 3D effects (and additionally due to the presence of
magnetic field and stellar activity/chromospheres) 
can be substantial for solar-type stars, but are not understood comprehensively at 
present\footnote{Approximate corrections for such 
effects may e.g. reproduce the upturn in the C/O ratio at low O as
indicated by the (non-LTE) data of \citet{fabbian09} also in the data of
\citet{spite05}. However, we prefer to display uncorrected values, as
these are likely more realistic in view of more recent investigations
(M. Spite, priv. comm.).}. The comparison of literature data for the
B-stars in the left panels and our data in the right panels shows once
more which improvements in the precision and accuracy of the analyses
were achieved. 

In terms of the investigation of the cosmic chemical evolution the current focus
of studies in the literature
lies on the early phases at low metallicity.
The interpretation of
the data comes from the comparison with Galactic chemical
evolution (GCE) models, which have to 
match the present-day composition 
as a boundary condition. It is therefore important for the entire
modelling which reference
values are used, solar or CAS abundances. In particular the
differences in the C/O ratio are appreciable, also with respect to the
majority of nearby solar-type stars, amounting to almost 50\%. Taken at face value 
this difference indicates that the
C/O enrichment of the ISM in the present-day solar neighbourhood occurred slower
than at the Sun's place of birth (see Sect.~\ref{sun-foreigner}).
No such conclusions could be drawn from the available literature data 
on early B-type stars so far.

Note that the Sun may be viewed as an extreme but still compatible
case in terms of the distribution of the stars from the present
sample in the log\,O--log\,C/O diagram, but the absolute values for
the carbon abundance differ significantly. 
The solar and CAS data on N/O are rather compatible on the other hand.
A systematic investigation of nitrogen abundances in high-metallicity
solar-type stars would be desirable for further comparisons.

The occurrence of $\alpha$-enhancement in the old stars of the halo and
of the thick disk, Fig.~\ref{mgfe_evolution}, is well understood in
terms of the different evolution timescales of supernovae of type Ia
and II. The CAS Mg/Fe-ratio is compatible with the data from the Sun
and from nearby solar-type stars, though it is somewhat low.
Metal-poor objects among the solar-type stars are easily explained by
their large lifetimes. However, metal-rich stars 
are absent among the young stars from the present sample, while they
are common among the solar-type stars. Finding a
reason for this is not straightforward in terms of standard 
GCE (metallicity is supposed to increase in time), but see the discussion 
in Sect.~\ref{sun-foreigner} related to radial migration. 
Again, data on early B-type stars from the literature were inconclusive 
in these terms.

Overall, it is astonishing how different and at the same time how
similar the young and old star populations in the solar neighbourhood
are. It is for the first time that this is elaborated, as the lack of
high precision and accuracy in many previous studies of early
B-type stars prevented any meaningful conclusions to be drawn.

In addition, reference abundances are not only of interest in
terms of the temporal evolution of elemental abundances, but also for
the spatial distribution, in particular for the interpretation of
Galactic abundance gradients. The current sample is not useful for
deriving abundance gradients per se because of the small baseline spread in
terms of Galactocentric radius (about 400\,pc). However, the CAS provides a
highly robust anchor point for theoretical models at the solar circle,
implying agreement with the solar standard for some elements, but 
also systematic shifts of various degrees for other elements.

\subsection{The Sun: an immigrant to its current neighbourhood}\label{sun-foreigner}
In a strict sense, comparisons of the CAS and the solar chemical composition
discussed in the previous sections are meaningful in terms
of GCE only when there is a causal
relationship of the 4.56\,Gyr old solar matter and the present-day 
material in its surroundings. In other words, the question is whether
the matter that we see in the nearby new-born stars, as represented 
in particular by the early B-type stars, and in the local ISM has
chemically evolved from the matter that was around at the birth of the
Sun. The existence of such a relationship is a strong assumption for most
previous and current GCE modelling\footnote{Technically, this is
realised in GCE models by a division of the Galactic disk into concentric 
annuli that evolve independently from each other. The question in our
context is, whether the Sun was born at a similar Galactocentric radius 
as it is observed today.}.

\begin{table*}[t!]
\footnotesize
\caption[]{Chemical tagging of the Sun's place of birth.\\[-6mm]\label{origin}}
\centering
\begin{tabular}{lccccccccc}
\hline\\[-2.5mm]
\hline
Element &  \multicolumn{2}{c}{Protosun} & & \multicolumn{2}{c}{Protosun, GCE corrected\tablefootmark{a}} & CAS & 
d$\varepsilon$(El.)/d$R_\mathrm{g}$ & \multicolumn{2}{c}{CAS+d$\varepsilon$(El.)/d$R_\mathrm{g}$} \\
\cline{2-3} \cline{5-6} \cline{9-10}
 & AGSS09 & CLSFB10 & & AGSS09 & CLSFB10 & & dex\,kpc$^{-1}$ & $R_\mathrm{g}$\,=\,6\,kpc & $R_\mathrm{g}$\,=\,5\,kpc\\[.5mm]
  \hline
C  & 8.47$\pm$0.05 & 8.54$\pm$0.06 & & 8.53$\pm$0.05 & 8.60$\pm$0.06 & 8.33$\pm$0.04 & $-$0.103$\pm$0.018\tablefootmark{b}  & 8.54$\pm$0.05 & 8.64$\pm$0.05\\
N  & 7.87$\pm$0.05 & 7.90$\pm$0.12 & & 7.95$\pm$0.05 & 8.01$\pm$0.12 & 7.79$\pm$0.04 & $-$0.085$\pm$0.020\tablefootmark{c}  & 7.96$\pm$0.05 & 8.05$\pm$0.05\\
O  & 8.73$\pm$0.05 & 8.80$\pm$0.07 & & 8.77$\pm$0.05 & 8.84$\pm$0.07 & 8.76$\pm$0.05 & $-$0.035\tablefootmark{d,e}          & 8.83$\pm$0.05 & 8.87$\pm$0.05\\
Mg & 7.64$\pm$0.04 & \ldots        & & 7.68$\pm$0.04 & \ldots        & 7.56$\pm$0.05 & $-$0.039\tablefootmark{d}            & 7.64$\pm$0.05 & 7.68$\pm$0.05\\
Si & 7.55$\pm$0.04 & \ldots        & & 7.63$\pm$0.04 & \ldots        & 7.50$\pm$0.05 & $-$0.045\tablefootmark{d}            & 7.59$\pm$0.05 & 7.64$\pm$0.05\\
Fe & 7.54$\pm$0.04 & 7.56$\pm$0.06 & & 7.68$\pm$0.04 & 7.70$\pm$0.06 & 7.52$\pm$0.03 & $-$0.052\tablefootmark{d}            & 7.62$\pm$0.03 & 7.68$\pm$0.03\\
\hline
\end{tabular}
\tablefoot{
\tablefoottext{a}{applying values from Table~5 of AGSS09, based on GCE models of \citet{CRM03}};
\tablefoottext{b}{\citet{esteban05}};
\tablefoottext{c}{\citet{carigi05}};
\tablefoottext{d}{\citet{cescutti07}, based on Cepheid observations of
\citet[and references therein]{andrievsky04}};
\tablefoottext{e}{a slightly steeper -- though compatible -- gradient,
by $-$0.044$\pm$0.010\,dex\,kpc$^{-1}$, is given by \citet{carigi05}.}
}
\end{table*}

Unfortunately, the birth-place of the Sun is unknown and tracing its
orbit back is a highly complex task. Passages near other stars or 
molecular clouds, which are a stochastic process, and dynamical interactions 
with spiral arms prevent a straightforward integration
of the Sun's orbit in the Galactic potential backward in time from
being successful. Improvements in our detailed understanding of the relevant
processes affecting the motions of stars will certainly result from 
the Gaia mission in the future. But for the moment, we can rely only on 
a statistical approach, using the theoretical framework of Galactic dynamics
and kinematics \citep[e.g.][]{wielen96,sellwood02} that predict that
old stars like the Sun are able to migrate up to several kiloparsecs
radially through the Galactic disk over their lifetime. It is therefore 
possible that the Sun was born away from the present solar circle. 
Chemical abundances can provide valuable additional constraints.

The radial migration of stars within the Galaxy has been 
incorporated in Galactic chemical evolution calculations for the first time 
by \citet{s09}. This kind of models constitutes a more realistic 
theoretical frame that could be used to refine the previous estimations  
\citep[e.g.][]{wielen96} of the solar place of origin. It is
important, though, to bear in mind that details in the model input like
e.g. the adoption of different yields or abundance gradients
along the Galactic disk may have a non-negligible effect in 
such estimations (R. Sch\"onrich, priv. comm.). The required systematic
studies are beyond the scope of the present paper, but we can provide
some qualitative evaluation, which is illustrated by
Table~\ref{origin}.

In order to make a meaningful comparison of metal abundances 
in terms of Galactic chemical 
evolution, CAS values need to be compared to the bulk composition of the Sun
(i.e. protosolar values\footnote{Over the lifetime of the Sun, 
the combined effects of thermal diffusion, gravitational settling and 
radiative acceleration have lead to a build-up of helium and metal 
abundances below the convection zone, such that the photospheric 
abundances are not representative for the bulk composition of the Sun.}), 
{\em corrected for the effects of GCE}. The
required abundance enrichments due to GCE are difficult to be quantified, 
as they depend on many model details like e.g. the adopted star-formation history 
and yields. Crucial is that enrichment is expected to occur, not depletion.
In view of the previous discussion in this section and in Sect.~\ref{GCE},
it is therefore in fact the similarities of the solar values
and the present-day CAS, which are astonishing, and less the differences.

Let us elaborate the argument in more detail: assume that the 
protosolar nebula would not have collapsed to the Protosun (with 
elemental abundances according to columns 2 and 3 of
Table~\ref{origin}). Instead, the gas would have been enriched in
metal content
over the past 4.56\,Gyrs as predicted by GCE models for the present 
solar vicinity, leading to the formation of the Sun at the
present day, with abundances according to columns 4 and 5. The Sun
would appear significantly more metal rich than its surroundings,
represented by the CAS (column 6).

A different birth place of the Sun than around the solar circle
could resolve this apparent contradiction.
Higher abundance values have been characteristic for 
the inner disk of the Milky Way for a long time over Galactic history.
This does not only apply to individual metal abundances, the argument
is further sustained by the presence of a higher C/O ratio in the Sun
than in the CAS (independent of the reference of solar abundances).
The inner disk shows a higher C/O ratio, which is supported both
by an observationally derived negative radial C/O gradient \citep{esteban05}
and some GCE models \citep[e.g.][]{carigi05}.

In order to further constrain the origin of the Sun, we have to
correct the CAS data for the effects of Galactic abundance 
gradients\footnote{Data are for the present-day, but GCE models suggest 
that the flattening of Galactic abundance gradients over the 
lifetime of the Sun are insignificant for our considerations 
\citep{marcon-uchida10}.} (column 7 in Table~\ref{origin}),
which were derived from a carefully analysed sample of \ion{H}{ii} regions
\citep{esteban05,carigi05} and the modelling of the Cepheid data of
\citet[and references therein]{andrievsky04} by \citet{cescutti07}.
From our own experience, we consider these works as sources of reliable data 
on this topic, see \citet{p08r} and \citet{firnstein10} for a discussion. 
The results from the gradient-corrected CAS values at Galactocentric 
radii $R_\mathrm{g}$ of 6 and 5\,kpc are shown in columns 8 and 9 of 
Table~\ref{origin}.

The comparison of the GCE-corrected protosolar abundances with the
gradient-corrected CAS values suggests that the birthplace of the Sun
was located at a $R_\mathrm{g}$ between 5 to 6\,kpc (depending on the
solar standard reference, which introduces another uncertainty to the previous
considerations). We conclude that the Sun -- and probably many 
other nearby older and metal-rich stars -- are immigrants to the present 
solar neighbourhood, supporting views that stellar radial migration 
is essential for understanding Galactic evolution \citep{s09}. 
It would be highly interesting though difficult to investigate the
details of the Sun's migration to its current location close to the 
Galactic co-rotation radius, which is so favourable for the existence 
of life on Earth \citep{leva98}.

\section{Summary\label{summary}}
With the present work we have established a new benchmark for analyses of
early B-type stars, demonstrating that practically the entire observed
spectra can be reproduced reliably. It was shown that by the combined use of
sophisticated models and of a thorough analysis methodology on
high-quality spectra {\em absolute} stellar parameters and elemental 
abundances for early-type stars can be determined with a precision
rivaling that of {\em differential} studies of solar-type stars. The
successful {\em simultaneous} match of many independent observational 
indicators like the Balmer line wings, multiple ionization equilibria,
SEDs and the agreement of spectroscopic and {\sc Hipparcos} distances
indicates that high accuracy was achieved at the same time, facilitated
by the minimisation of systematic errors wherever they could be identified. 

Overall, our analysis methodology provides both the accuracy and precision 
to use early B-type stars as versatile tools for astrophysics, besides
our immediate objective here. Some conclusions from earlier applications 
get even strengthened in retrospect by the present work: e.g. on chemical tagging 
in order to determine population membership of hypervelocity stars 
\citep{p08a,tillich09}, the investigation of subtle abundance signatures to 
constrain supernova nucleosynthesis from binary supernova runaways
\citep{p08b,i10}, or on coupling quantitative spectroscopy with
asteroseismic analysis of pulsating OB stars observed with {\sc CoRoT} 
\citep{b10}. And, the methodology offers a high potential for future applications,
that may facilitate many more facets of astrophysics to be studied at high 
confidence.

The principal application of our novel analysis methodology here was on
a larger sample of nearby apparently slowly-rotating early B-type stars 
that were cleaned from peculiar 
objects. This confirmed that the young stellar component of the 
solar neighbourhood is chemically homogeneous to better than
10\%, in accordance to studies of the local ISM abundances -- the
material out of which the stars formed. This in turn allowed us 
to establish a present-day cosmic abundance standard (CAS), which has the 
advantage of redundancy when compared to the solar standard, as it is
based on an entire sample of stars instead of one object alone.
So far, information for the most abundant chemical species have been
provided, with $\log \varepsilon$\,$\ge$\,7.50.

First implications of the existence of a present-day cosmic abundance
standard were outlined here. The CAS represents the recommended 
initial chemical composition for stellar structure and evolution
calculations, in particular for short-lived massive stars.
Observationally, nitrogen enrichment in about 1/3 of the sample 
objects indicates mixing of the surface layers with CN-processed
material from the stellar core, with the abundance ratios
for the light elements carbon, nitrogen and oxygen following tightly the
predicted nuclear path of the CNO cycles.

The CAS provides the to date most authoritative reference for 
constraining the chemical composition of the ISM dust-phase. A
silicate-rich and relatively carbon-poor nature of the local ISM dust
is inferred, challenging many contemporary dust models. In combination
with the finding that carbonaceous dust is practically absent inside
the Orion \ion{H}{ii} region this implies that the dust in the ISM
is in its majority composed of silicates, PAHs, organic refractory
material and possibly amorphous carbon, with only little carbon incorporated 
in graphite.

The CAS provides tight reference points for anchoring models of the chemical 
evolution of the Milky Way, constraining the present-day endpoint of
Galactic nucleosynthesis in the solar neighbourhood in the course of
the cosmic matter cycle. We discussed in particular the examples 
of the light elements CNO, and the Mg/Fe ratio as a tracer for the
$\alpha$/Fe ratio. While the CAS indicates a surprisingly good agreement 
with most inferences from the solar abundances, a striking difference is found 
for the C/O ratio, amounting to almost 50\% between cosmic and solar 
values. Intriguingly, early B-type stars with super-solar metallicity
are absent in the solar neighbourhood, while several nearby solar-type
stars show super-solar metallicities. This can be interpreted as a
consequence of radial migration of stars within the Galactic disk.

Finally, the comparison of the CAS with the solar standard
in view of Galactic chemical evolution (GCE) shows that the highly
successful use of the Sun as proxy for cosmic abundances is somewhat
coincidental. Radial migration outward from its birth place in the
inner disk at $\sim$5-6\,kpc Galactocentic distance (where higher
metallicity values were reached earlier in cosmic history) over its
lifetime to its current neighbourhood has compensated for the expected metal 
enrichment in the course of GCE. A telltaling signature is left only 
in the C/O ratio. The present work provides the so far most 
stringent evidence in terms of chemical signature that the Sun --
like many other solar-type stars -- is
an immigrant to its current Galactic neighbourhood.

\begin{acknowledgements}
We wish to thank M.~Firnstein and T.~Kupfer for help with the reduction of 
{\sc Foces} spectra, K.~Butler for providing {\sc Detail/Surface}, 
H.~Hirsch for making {\sc Spas} available and E.~Niemczura for 
computations of {\sc Atlas9} models. 
We are grateful to M. Asplund, U. Heber, 
G. Meynet, R. Sch\"onrich, S. Sim\'on-D\'iaz, E. Ntormousi
and the MPA SAGA-group for fruitful discussions, and the referee for valuable
suggestions. We thank the staff at Calar Alto 
for performing observations in the service run of 2005. 
Some of the data presented in this paper were obtained from
the Multimission Archive at the Space Telescope Science
Institute (MAST). STScI is operated by the Association of
Universities for Research in Astronomy, Inc., under NASA
contract NAS5-26555. Support for MAST for non-HST data is
provided by the NASA Office of Space Science via grant NAG5-7584
and by other grants and contracts. This research has made use of
the SIMBAD database, operated at CDS, Strasbourg, France.
Travel to the Calar Alto Observatory/Spain in 2001 was
supported by \emph{DFG} under grant PR\,685/1-1.
\end{acknowledgements}

\renewcommand{\thefigure}{\arabic{figure}}
\clearpage
\end{document}